\documentclass[11pt,hyper,letterpaper]{JHEP3}
\usepackage{verbatim}
\usepackage{amssymb,amsfonts,amsthm,amsmath,amssymb,amscd,amstext,epsfig}

\newcommand{\bea}{\begin{eqnarray}}
\newcommand{\eea}{\end{eqnarray}}

\newcommand{\II}{\mathcal{I}}
\newcommand{\JJ}{\mathcal{J}}

\def\be{\begin{equation}}
\def\ee{\end{equation}}

\def\ut{\underline{t}}
\def\ux{\underline{x}}

\def\ur{\underline{r}}

\def\unu{\underline{\nu}}
\def\ulambda{\underline{\lambda}}
\newcommand{\du}[2]{_{#1}^{\phantom{#1}#2}}
\newcommand{\twobyone}[2]{\left(\begin{array}{c} #1 \\ #2 \end{array}\right)}
\newcommand{\twobytwo}[4]{\left(\begin{array}{cc} #1 & #2 \\ #3 & #4 \end{array}\right)}

\preprint{PUPT-2352
}
\title{Sum Rules from an Extra Dimension}
\author{
Daniel R.~Gulotta\footnotemark[1]\,,
Christopher P.~Herzog\footnotemark[2]\,, 
and Matthias Kaminski\footnotemark[3]\\
Department of Physics, Princeton University \\
     Princeton, NJ 08544, USA \\
}

\footnotetext[1]{\email{dgulotta@princeton.edu}}
\footnotetext[2]{\email{cpherzog@princeton.edu}}
\footnotetext[3]{\email{mkaminsk@princeton.edu}}

\abstract{
Using the gravity side of the AdS/CFT correspondence, we investigate the analytic properties of thermal retarded Green's functions for scalars, conserved currents, the stress tensor, and massless fermions.  We provide some results concerning their large and small frequency behavior and their pole structure.  From these results, it is straightforward to prove the validity of various sum rules on the field theory side of the duality.
We introduce a novel contraction mapping we use to study the large frequency behavior of the Green's functions.
}
\date{November 2010}
\begin{document}
\setcounter{page}{1}
\setcounter{equation}{0}

\section{Introduction}

The AdS/CFT correspondence (or more generally gauge/gravity dualities) provide a recipe for computing correlation functions for strongly interacting field theories from classical gravity \cite{Maldacena:1997re,Gubser:1998bc,Witten:1998qj}.
The analytic properties of field theory Green's functions are constrained by considerations such as unitarity and causality.  For example, causality forces the Fourier transform of the retarded Green's function to be analytic in the upper half of the complex frequency plane (with our sign conventions).  
If the AdS/CFT conjecture is true, these analytic properties must hold. However, without a proof of the conjecture it is an interesting question to
ask how these analytic properties arise from the gravitational side of the correspondence.  In particular, we would like to understand how sum rules for thermal Green's functions 
emerge from gravity.

Sum rules play an important if sometimes overlooked role in non-perturbative field theory.
There has been recent interest in applying sum rules to investigate the strongly interacting quark-gluon plasma formed at the Relativistic Heavy Ion Collider (RHIC). To compute transport coefficients from lattice QCD calculations, Euclidean Green's functions need to be continued to real time.  In this difficult game, sum rules provide additional constraints that the analytically continued Green's functions must satisfy.  To obtain bulk and shear viscosities from the stress tensor two-point function, sum rule constraints were employed by
for example refs.\ \cite{Kharzeev:2007wb,Karsch:2007jc,Meyer:2010ii,Meyer:2010gu}.
One may also consider sum rules for a conserved current.  One famous result is the Ferrell-Glover-Tinkham sum rule which relates the ``missing area'' in the real part of the conductivity as a function of frequency to the London penetration depth 
of a BCS 
superconductor.\footnote{%
Recall there exists a Kubo relation between the current-current correlation function and the conductivity.
The London penetration depth is related to the superfluid density which in turn is proportional to the ``missing area''.
}

In the AdS/CFT context, there have been already a number of papers about thermal sum rules.
Romatschke and Son \cite{Romatschke:2009ng} derived a pair of sum rules for the
 stress tensor of 
${\mathcal N}=4$ SU($N$) Super Yang-Mills theory in 3+1 dimensions at large $N$ and strong coupling. 
Similar sum rules for a non-conformal theory dual to a Chamblin-Reall background were constructed in \cite{Springer:2010mf, Springer:2010mw}. 
The authors of \cite{Hartnoll:2008vx} numerically demonstrated an analog of the Ferrell-Glover-Tinkham sum rule for a holographic superconductor in 2+1 dimensions, while \cite{Baier:2009zy} studied an R-charge correlator sum rule for ${\mathcal N}=4$ Super Yang-Mills theory.
In this paper, we will rederive the results of \cite{Romatschke:2009ng,Hartnoll:2008vx,Baier:2009zy} in a slightly more rigorous and general framework.  Our point of view is different; we want to see how gravity constraints enforce the sum rules more generally rather than checking in specific cases numerically or analytically that the sum rules are valid.

In Section \ref{sec:holoTheories}, we present the 
class of gravity theories that we intend to analyze and summarize our results.  The bosonic correlation functions we study are governed through the gravity theory by a single second order differential equation.  Instead of considering scalar, current, and stress-tensor correlators separately, we unify the analysis through studying the behavior of this single differential equation.  In contrast, for the massless fermions, we study only correlators of objects with spin 1/2.

In Section \ref{sec:sumrules} we review the 
conditions which lead to sum rules in field theories.
Taking the corresponding gravity point of view in Sections \ref{sec:gravity} and \ref{sec:fermions}, we state our assumptions and 
derive sum rules from gravity. In particular, we show that -- given our assumptions -- the 
Green's functions are holomorphic in the upper half of the complex frequency plane including the real axis but not the origin.  

Furthermore, we
derive a contraction map for bosons and a contraction-like map for massless fermions 
 which we use to bound
the large frequency behavior of the corresponding spectral functions.
From these two major results we derive general sum rules in
various dimensions and relate them to earlier results. 
Finally, we discuss our work and suggest some
future directions in Section \ref{sec:discussion}.

\section{A Class of Holographic Theories \& Summary of Results}
\label{sec:holoTheories}
Before we describe our class of theories, we present our field theory definition of the retarded Green's function for 
two bosonic operators\footnote{Note that our definition of the Green's function differs from the one given in 
\cite{Son:2002sd} by a minus sign.}
\be
\tilde G_R^{ij}(t,x) \equiv i \theta(t) \langle [ {\mathcal O}^i(t,x), {\mathcal O}^j(0) ] \rangle \ ,
\ee
and two fermionic operators
\be
\tilde G_R(t,x) \equiv i \theta(t) \langle \{ {\mathcal O}(t,x), {\mathcal O}^\dagger(0) \} \rangle \ .
\ee
Our conventions for the Fourier transform are
\be
G_R(\omega,k) \equiv \int d^dx \, e^{i \omega t - i k x} \tilde G_R(t,x) \ ,
\ee
where we have taken advantage of translation invariance of the underlying theory.  
We define also the spectral density to be the imaginary part of $G_R(\omega,k)$:
\be
\rho(\omega, k) \equiv \frac{1}{2i} \left( G_R(\omega, k) - G_R(\omega, k)^\dagger \right) \ .
\label{specdensity}
\ee

On the gravity side we will not be completely general and instead focus on a particular class of $d+1$ dimensional space-times with the line element:
\be
\label{eq:metric0}
ds^2 = - f(r) e^{-\chi(r)} dt^2 + \frac{r^2}{L^2} \, d \vec x^2  + \frac{dr^2}{f(r)}  \ .
\ee
We will assume that the space-time contains a horizon at $r=r_h$ where $f(r_h) = 0$ and 
$f'(r_h) = f_h \neq 0$.  We also assume that at large $r$, the space-time becomes asymptotically anti-de Sitter with a radius of curvature $L$: 
\be
f = \frac{r^2}{L^2} + \ldots \ , \; \; \; e^{-\chi} f = \frac{r^2}{L^2} + \ldots \ .
\ee
These space-times are conjectured to be dual to thermal field theories with $d$ space-time dimensions and have appeared many times before in the AdS/CFT literature.  They are sufficiently general to allow us to consider field theories at nonzero temperature and charge density and also, if need be, in a superconducting or superfluid phase 
\cite{Gubser:2008px, Hartnoll:2008kx, Herzog:2008he}.

The prescription for calculating Green's functions was provided in the original papers \cite{Gubser:1998bc,Witten:1998qj} and later extended to nonzero temperature by
\cite{Son:2002sd,Herzog:2002pc}.  
The procedure begins by finding a solution to a system of differential equations describing fluctuations of classical gravitational fields.  
In this paper, we will consider the two-point functions for a scalar, a conserved current, the stress tensor, and a spin 1/2 field in the space-time (\ref{eq:metric0}).
  
For bosonic fields, 
the following versatile second order differential equation will play a big role in this paper:
\be  \label{eq:order2}
{\mathcal D}(\phi)(r) \equiv \phi''(r) + T_1(r) \phi'(r) + T_0(r) \phi(r) = 0 \ ,
\ee
where $'$ denotes $d/dr$, 
\be
T_1 = \frac{F'}{F} + \frac{n}{r} \qquad \mbox{and} \qquad T_0 = \frac{\omega^2}{F^2} - \frac{Y}{F} \ ,
\ee
$Y$ is a potential term specified below, and $F = f e^{-\chi/2}$. 
This differential equation can be used to study two-point functions of scalar, vector and tensor fields from gravity.  

After a brief review of sum rules in Section \ref{sec:sumrules}, in Section \ref{sec:gravity} we will study the analytic structure of the Green's functions for a scalar ${\mathcal O}$, a component of a conserved current $J^x$, and a component of the stress-tensor $T^{xy}$ using the differential equation (\ref{eq:order2}).  
We discuss the reality and positivity properties of $G_R(\omega,k)$:  $G_R(\omega,k)^* = G_R(-\omega, -k)$ and $\omega \, \rho(\omega, k) \geq 0$.
We show that $G_R(\omega,k)$ continued to complex $\omega$, has no poles or 
branch cuts in the upper half of the complex plane including real $\omega \neq 0$. 
We discuss the large $\omega$ behavior of $G_R(\omega,0)$ from which we can justify  the sum rules. 
Finally, we repeat this analysis for massless fermions in Section \ref{sec:fermions}.

Let us first explain how (\ref{eq:order2}) governs the bosonic correlators and describe the results for fermions and bosons in greater detail:

\subsection*{Scalars}
Consider a real scalar operator ${\mathcal O}$ dual to a minimally coupled scalar field $\phi$ in the gravity dual:
\be
S_\phi = -\int d^{d+1} x \sqrt{-g} \left[ (\partial_\mu \phi) (\partial^\mu \phi) + m^2 \phi^2 \right] \ .
\ee
For $n=d-1$ and $Y = (m^2 + k^2 L^2 / r^2) e^{-\chi/2}$ 
the differential equation (\ref{eq:order2}) is the equation of motion of a scalar field $\phi$ of mass $m$ and space-time dependence $\phi \sim e^{-i \omega t + i k x}$.  
The momentum space two-point function for ${\mathcal O}$ can be read off from the large $r$ expansion of $\phi$.  The expansion takes the form
\be
\phi = a r^{\Delta-d} (1 + \ldots) + b r^{-\Delta} (1 + \ldots ) \ ,
\ee
where the ellipsis denotes higher order terms in $r$ and $\Delta(\Delta-d) = m^2 L^2$.  Note that $\Delta$ can be interpreted as the scaling dimension of the dual operator in the field theory.  
From gravity, solving the differential equation with ingoing boundary conditions at the event horizon $r=r_h$, up to an overall normalization and possible logarithmic contact terms, we have
$G_R(\omega, k) \sim -b / a$.  

We do not press the analysis as far as a demonstration of sum rules, with the exception of $m=0$, where the analysis overlaps with the stress tensor correlator described below.

\subsection*{Conserved Currents}
Conserved currents $J^\mu$ in field theory are dual to gauge fields $A_\mu$ in the gravity dual.  
If we set $n=d-3$, $\phi = A_x$, and $Y$ to a particular form we describe in Appendix \ref{app:conductivity} (see (\ref{Axeq})), then a solution to (\ref{eq:order2}) describes the $G^{xx}(\omega,0)$ component of the current-current Green's function in the dual field theory.  Indeed, the presence of $Y$ allows us to study $G_R$ as a function of charge density and also follow its behavior through a superfluid phase transition.

The Green's function can be found from the behavior of the
gauge field near the boundary.  Restricting to $d=3$ and 4, the $x$-component of the gauge
field takes the form $A_x = A_x^{(0)}+A_x^{(1)} r^{-(d-2)} + O(r^{-(d-1)})$. 
Being more careful about normalization than we were in the scalar case, 
the Green's function is then (see for example \cite{Son:2002sd})
\begin{equation}
G_R^{xx}(\omega,0)=  \frac{d-2}{e^2 L^{d-1}}\frac{A_x^{(1)}}{A_x^{(0)}} = 
\frac{-1}{e^2 L^{d-3}} \lim_{r \to \infty}  \frac{F r^{d-3} A_x'}{A_x} \ ,
\label{GRxxdef}
\end{equation}
where $e$ is a coupling constant for the $U(1)$ gauge field strength.
In $d=4$, this expression will contain a logarithmically divergent contact term that needs to be regularized.

In the cases $d=2$, 3 and 4 when 
$Y = \chi=0$, and $f = r^2 ( 1 - (r_h/r)^d) / L^2$, the equation 
(\ref{eq:order2}) is exactly solvable.  
This case corresponds to a thermal field theory at zero charge density (and also not in a superfluid or superconducting phase).
One can explicitly verify the analytic structure of the Green's function.  We review these known results in Appendix \ref{app:JJ}.

The more interesting general case where $Y \neq 0$ we save for Section \ref{sec:gravity}.  
Given some assumptions about the behavior of $G_R^{xx}(0,0)$, we are able to prove that the Ferrell-Glover-Tinkham sum rule holds from gravity considerations.

\subsection*{Stress Tensor}
The stress tensor $T^{\mu\nu}$ in field theory is dual to fluctuations of the metric in the gravity dual.  It is well known that off-diagonal metric fluctuations $\delta g_{xy} = \phi(r) e^{-i \omega t + i k z} r^2 / L^2$ that are not functionally dependent on $x$ or $y$ obey the equation for a minimally coupled scalar of mass $m=0$.  Thus, a solution to (\ref{eq:order2}) for $n=d-1$ allows a calculation of a particular component of the stress-tensor two point function: $G_R^{xy,xy}(\omega, \vec k)$ where the $x$ and $y$ components of $\vec k$ are set to zero. 
Up to contact terms and regulators we describe in more detail in Appendix \ref{app:TT}, the retarded Green's function is
\be
G_R^{xy,xy}(\omega,k) =  \frac{-1}{2 \kappa^2 L^{d-1}} \lim_{r \to \infty} \frac{F r^{d-1} \phi'}{ \phi} \ ,
\ee
where $\kappa$ is the gravitational coupling constant.
This correlator is of particular interest since
it is related to the shear viscosity transport coefficient~\cite{Policastro:2002se}.
Among other results, we recover in general space-time dimension a sum rule proposed by Romatschke and Son \cite{Romatschke:2009ng} for ${\mathcal N}=4$ Super Yang Mills, i.e. the $d=4$ case.
For the sum rule, we restrict to $Y=0$ and hence $k=0$.  Note that $k=0$ is still the interesting case for the shear viscosity computation.

\subsection*{Fermions} The operators ${\mathcal O}$, $J^x$, and $T^{xy}$ are all Hermitian.  By way of counterpoint, in Section \ref{sec:fermions}, we consider sum rules for a non-Hermitian operator --- a minimally coupled massless fermion in an asymptotically $AdS_{d+1}$ space. The relation $G_R^{\alpha \beta}(\omega,k)^*= G_R^{\alpha \beta}(-\omega, -k)$ (with $\alpha, \beta$ spinor indices)
does not hold.
We show $\rho (\omega, k ) >0$ for real $\omega \neq 0$.\footnote{%
The missing factor of $\omega$ in this inequality can be traced to the different statistical distributions obeyed by bosons and fermions.  The Fourier transform of the variance, 
 $\langle {\mathcal O}^\dagger (x) {\mathcal O}(0) \rangle$, is non-negative, as can be seen by decomposing ${\mathcal O}(x)$ into a complete basis of eigenstates. The variance is then related to the spectral density by a Bose-Einstein or Fermi-Dirac distribution function, depending on whether ${\mathcal O}(x)$ is a bosonic or fermionic operator.
 }
 We prove that all (quasi)normal modes must either lie in the lower half of the complex frequency plane, or 
at $\omega =0$.  
Finally, we study the large $\omega$ behavior of $G_R(\omega,k)$. 
These are the ingredients from which we derive a sum rule for massless, charged 
fermions in arbitrary $d$.
For $d=3$, the sum rule we find is a special case of a sum rule proposed in an 
Appendix of \cite{Benini:2010pr}.

\section{Sum Rules from Field Theory}
\label{sec:sumrules}

Consider a retarded Green's function $G_R(\omega)$ and the corresponding spectral density,  $\rho(\omega)$ (\ref{specdensity}).
 While this section can be generalized to the case where $G_R(\omega)$ has a multi-index structure, for clarity we will assume that $G_R(\omega)$ is the single component retarded Green's function of a scalar bosonic operator.
We make two assumptions about the retarded Green's function:
\begin{enumerate}
\item
$G_R(\omega)$ is holomorphic in the upper half plane, including the real axis. 
\item
$
\lim_{|\omega| \to \infty} G_R(\omega) = 0 \ 
$
if $\mbox{Im} \, \omega \geq 0$.
\end{enumerate}

The first assumption enables us to apply the Cauchy integral theorem:
\begin{equation}
G_R (\omega+i \epsilon) = \oint_C \frac{G_R(z) dz}{2\pi i(z-\omega - i \epsilon)} \; ; \; \; \;
0 = \oint_C \frac{G_R(z) dz}{2\pi i(z-\omega + i \epsilon)} \ .
\end{equation}
For $\omega \in \mathbb{R}$ and some $r \in \mathbb{R}^+$, 
we choose $C$ to be the contour that
travels along the real axis from $-r$ to $r$ and then along a semicircle
in the upper half plane back to $-r$.
Our second assumption tells us that the integral along the
semicircle must go to zero as $r \rightarrow \infty$.  Then we have
\begin{equation} \label{eq:cauchy}
G_R(\omega+i \epsilon) = \lim_{r \rightarrow \infty} \int_{-r}^{r}
\frac{G_R(z) dz}{2\pi i(z-\omega - i \epsilon)} \; ; \; \; \;
0= \lim_{r \rightarrow \infty} \int_{-r}^{r}
\frac{G_R(z) dz}{2\pi i(z-\omega + i \epsilon)} \ .
\end{equation}
We subtract the complex conjugate of the second integral from the first integral in
\eqref{eq:cauchy}
and take the limit $\epsilon \rightarrow 0$ to find
a ``spectral representation'' of the retarded Green's function:
\be
G_R(\omega) = \lim_{\epsilon \to 0^+}  \int \frac{d z}{\pi} \frac{\rho(z)}{z - \omega - i \epsilon} \ .
\label{specrep}
\ee

If assumption 2 is not satisfied, then a modified version of (\ref{specrep}) can often be constructed,
\be
\delta G_R(\omega) = \lim_{\epsilon \to 0^+}  \int \frac{d z}{\pi} \frac{\delta \rho(z)}{z - \omega - i \epsilon} \ ,
\label{deltaspecrep}
\ee
where  $\delta \rho = \mbox{Im} \, \delta G_R$. 
The precise definition of $\delta G_R(\omega)$ depends on the regularization procedure.
For example, suppose
\be
G_R(\omega) = \delta G_R(\omega) + c_0 + c_1 \omega + \ldots + c_n \omega^n 
+( d_0 + d_1 \omega \ldots +d_s \omega^s) \log(-i \omega)
\ee
where $\delta G_R(\omega)$ does satisfy assumption 2.
We take the branch cut of the log to lie along the negative imaginary axis.
The relation (\ref{deltaspecrep}) then replaces (\ref{specrep}).
Other regularizations are often possible.  
The coefficients $c_i$ and $d_j$ may be independent of a parameter in the theory, for example temperature $T$.  In this case, we could construct a regulated $\delta G_R$ by considering the difference  $G_R(T_1) - G_R(T_2)$.
Another way to regulate the large $\omega$ divergence is to take derivatives:
\be
\frac{\partial^{n+1}}{\partial \omega^{n+1}} G_R(\omega) =   \lim_{\epsilon \to 0^+}\int \frac{dz}{\pi} \frac{(n+1)! \, \rho(z)}{(z-\omega-i \epsilon)^{n+2}} \ . 
\ee
While the entire function $G_R(\omega)$ is generically 
difficult to determine in an arbitrary quantum field theory, it is often possible to obtain the $c_i$ and $d_j$ in the limit $\mbox{Im} \, \omega \to \infty$, for example through an operator product expansion in an asymptotically free theory.

Given an appropriately regulated version of (\ref{specrep}), people often call the spectral representation evaluated at $\omega=0$ a sum rule:
\be
\delta G_R(0) = \int \frac{d\omega}{\pi} \frac{\delta \rho(\omega)}{\omega} \ .
\label{genericsumrule}
\ee
By construction, this object is convergent for large $\omega$, but in discarding the $i\epsilon$ regulator, we have to be careful about the convergence properties of the integral at 
$\omega=0$.   
From the definition of the retarded Green's function,
it follows for bosonic Hermitian operators that $G_R(\omega, k)^* = G_R(-\omega,-k)$ (for real $\omega$ and $k$).  Let us restrict to the case $k=0$.  
In a small $\omega$ expansion of $G_R(\omega,0)$, even powers of $\omega$ have real coefficients and odd powers have pure imaginary coefficients.  Provided $\delta G_R(0,0)$ is finite, the sum rule should be well defined.\footnote{%
 We will see a case in the appendix, for the current-current correlation function in $d=2$ space-time dimensions, where $G_R(\omega) \sim i/\omega$ at small frequencies.  In this case, we have to be more creative to write a sum rule.  Note that this behavior violates assumption 1 above.
 \label{footnote2}
}

There are
also generalized sum rules that involve derivatives of $G_R(\omega)$:
\be
\frac{1}{(2j)!} \frac{\partial^{2j}}{\partial \omega^{2j}} G_R(0) =   
\int \frac{d\omega}{\pi \omega^{2j+1}} \left( \rho(\omega)  - \sum_{i=1}^j a_{2i-1} \omega^{2i-1} \right) \ , 
\label{gensumrule}
\ee
where $2j > n,s$.  For these generalized sum rules, the $a_i$ are chosen to ensure that the integral converges at $\omega=0$.  Note we have taken advantage of the reality properties of a small $\omega$ expansion of $G_R(\omega)$.  
Some specific examples of this type of generalized sum rule for current-current correlators are given in the appendix, (\ref{d2gensum}) and (\ref{d4gensum}).

\section{Sum Rules from Gravity}
\label{sec:gravity}

We would like to see how correlation functions for bosonic Hermitian operators derived from the gravity side of the AdS/CFT correspondence satisfy sum rules of the form (\ref{genericsumrule}).  
(The fermionic case is postponed to Section \ref{sec:fermions}.)
Consider a generic retarded Green's function $G_R(\omega,k)$.  We will establish, from gravity, that $G_R(\omega,k)$ is holomorphic in the upper half plane and also for real $\omega \neq 0$.  
We will check that $G_R(\omega,k)^* = G_R(-\omega,-k)$ and that
$\omega \rho(\omega, k) \geq 0$ (for real $\omega$ and $k$).
We will determine the large frequency behavior of $G_R(\omega, 0)$ 
so that we can appropriately regularize (\ref{specrep}). 
We will not be able to show in general that $G_R(0,0) < \infty$.\footnote{%
See footnote \ref{footnote2}.
} 

\subsection{$\mathrm{Im}\,\omega>0$}
\label{sec:omegaGreater0}
There is a very good, well known physical reason to expect that $G_R(\omega)$ is holomorphic in the upper half plane.  The reason is field theoretic and related to causality.  Consider the Fourier transform back to real space
\be
\tilde G_R(t) = \int \frac{d \omega}{2\pi} e^{-i \omega t} G_R(\omega) \ .
\ee
If $G_R(\omega)$ has poles and branch cuts only in the lower half plane, then the contour can be closed in the upper half plane for $t<0$ and the integral evaluates to zero, as expected if the response of the system is to be causal.\footnote{
Note that we could redefine our Green's function such that the integration contour passes above all of the poles.  While the resulting Green's function 
will be causal, the contour deformation is non-standard, and the poles in the upper half plane indicate instabilities in the field theory. 
We thank Karl Landsteiner for discussion on this point.}

As emphasized in the introduction, we would also like to have a gravitational reason.
In Appendix \ref{app:branchCuts} we show that there are no branch cuts in the class of theories we analyze here. So we consider only meromorphic $G_R(\omega, k)$, and the only non-analyticities are poles which ref.\ \cite{Son:2002sd} has argued are dual to quasinormal modes in gravity. 
The locations $\omega^{\mathrm{pole}}_n$ of poles of the boundary theory Green's function in the 
complex frequency plane are exactly the
quasinormal frequencies $\omega^{\mathrm{qnm}}_n$ on the gravity side.
A quasinormal mode is a solution where the 
bulk field has ingoing boundary conditions at the event horizon of the black hole and where
the
leading behavior of the field
vanishes, e.g.\ $A_x^{(0)}=0$ for the current-current correlation function.
Given the assumed time dependence $e^{-i \omega t}$, a quasinormal mode in the upper half of the complex plane is a solution that is exponentially growing in time, indicating an instability of the metric and/or matter fields in the gravity dual.
  An interesting consequence of gauge/gravity duality is thus that 
 the boundary requirement of causality $\mathrm{Im}(\omega^{\mathrm{pole}}_n)\le 0$
corresponds to the gravity requirement of stability $\mathrm{Im}(\omega^{\mathrm{qnm}}_n)\le 0$.\footnote{
Causality can also be studied considering the front velocity, and analyzing the poles of $G_R(\omega,k)$ in the {\it complex momentum} plane (for real frequencies)~\cite{Amado:2007pv}.}

Just because a pole in the upper half of the complex plane would be a bad thing doesn't mean it can't happen.  (Indeed, such poles drive the holographic superconducting phase transition described in \cite{Hartnoll:2008vx,Gubser:2008px}). Given some additional mild technical assumptions, 
we now prove for the class of scalar, current, and stress tensor correlation functions governed by the differential equation (\ref{eq:order2}) introduced above, 
there are poles only in the lower half of the complex plane and at $\omega = 0$.
In particular, we make some general assumptions about (\ref{eq:order2}).
We assume that $n \geq 0$.  
Next, we assume $F>0$ and $Y \geq 0$ on the interval $r_h < r< \infty$.  We also assume the near horizon behavior $F \approx (r-r_h)F_h>0$ and $Y(r_h) < \infty$ and the large $r$ behavior
$F \sim r^2$ and $Y \sim r^{-2 \Delta}$ where $\Delta >0$.
From these assumptions, it follows that 
a quasinormal mode solution to this differential equation has the near boundary 
behavior\footnote{%
If $\Delta=0$, as for example happens for a massive scalar field, we have $Y\sim m^2$ near the boundary.  Assuming $m^2>0$, it follows that $A$ has 
an even smaller exponent $A \sim r^{-1/2(n+1+\sqrt{(n+1)^2+4 m^2 L^2 })}$ at large $r$.  The argument we are about to present justifying the absence of quasinormal modes in the upper half plane continues to hold.
}
 $A \sim r^{-n-1}$ and the near horizon behavior 
\be
A \sim (r-r_h)^{-i \omega /F_h} \ .
\label{nearhorizonA}
\ee
The near horizon behavior is chosen to give a retarded rather than an advanced Green's function.  With an implicit time dependence $e^{-i \omega t}$, the boundary condition corresponds to a plane wave traveling into the event horizon.

Such a differential equation (\ref{eq:order2}) 
can be derived from the one dimensional effective action
\be
S_A = \int_{r_h}^\infty dr \, F r^n \left[ |A'(r)|^2  + \left(\frac{Y}{F}- \frac{\omega^2}{F^2}\right) |A(r)|^2 \right] \ .
\label{SAaction}
\ee
Assume that there exists a quasinormal mode solution $A$ to (\ref{eq:order2}) 
with frequency $\omega$.  
Because the differential operator is real, there will be a second complex conjugate quasinormal 
mode $\bar A$ with frequency $\bar \omega$.
Consider $0=S_A-S_A$, evaluated on implicit
solutions $A$ and $\bar A$. We integrate by parts, 
using the equation of motion for $\bar A$
in the first $S_A$, and that for $A$ in the second. We find that
\be
0=\left. F r^n \left[ \bar A' A - \bar A A' \right] \right|_{r_h}^\infty + (\bar \omega^{2} - \omega^2) \int_{r_h}^\infty  dr \, \frac{r^n}{F} |A|^2 \ .
\label{intparts}
\ee
Because of the large $r$ behavior of $A$, the first term must vanish evaluated as $r \to \infty$ for $n \geq 0$.
Now if $\omega$ lies in the upper half of the complex plane, 
the first term will also vanish evaluated at $r=r_h$.  In this case, $\bar \omega^2 = \omega^2$
and we find that $\omega$ lies on the positive imaginary axis.
For $\omega$ in the lower half plane, in contrast, the differential operator is not self-adjoint and 
$\omega^2$ need not be real.

Assume now that there exists a quasinormal mode in the upper half plane 
for which $\omega^2 < 0$.  In this case, $S_A$
is positive definite.  If we integrate by parts and use the equation of motion (\ref{eq:order2}), $S_A$ reduces to the boundary term
\be
S_A = \left. F r^n \bar A(r) A'(r) \right|_{r_h}^\infty \ .
\ee
From the near boundary and near horizon behavior of $F$ and $A$, $S_A = 0$.  Thus, no quasinormal modes in the upper half of the complex plane exist.  

Note if we weaken the condition on $Y$ and allow $Y$ to become negative and develop a potential well, then such a quasinormal mode can exist. 
As an example of such a quasinormal mode crossing into the upper half plane, consider the case where $A$ is a {\it charged} scalar field in the holographic superconductors
\cite{Hartnoll:2008vx,Gubser:2008px,Hartnoll:2008kx,Amado:2009ts}. Because the scalar is charged, it does not fall in the class of fluctuations which we address in this paper, and its
instability is thus not ruled out by our argument.
As a similar example, consider the holographic
superconductor involving instability of a {\it neutral} scalar~\cite{Hartnoll:2008kx}. Here a negative
mass squared causes $Y<0$ violating our initial assumption $Y\ge0$, and this case again is not in our class of models.\footnote{%
We suspect that $Y<0$ also for a  third example of instability: 
the {\it real} scalar fluctuation of the probe brane embedding in the 
D3/D7 system~\cite{Kaminski:2009ce}. Based on numerical evidence, we know the Schr{\"odinger} potential of the scalar develops a well. 
}
In both cases, this scalar instability causes the phase transition to the superconducting state. In the new stable phase, the scalar develops a nonzero expectation value. 
(The field which we call $\Phi$ 
later in this paper could be thought of as such a condensate.)

\subsection{Real $\omega$} \label{sec:realomega}

The observant reader will have noted that while both gravity arguments above rule out 
quasinormal modes with $\mbox{Im} \,\omega >0$, they do not rule out quasinormal modes 
with $\mbox{Im} \, \omega=0$.  In Section \ref{sec:sumrules}, 
however, we required the Green's function to be holomorphic also 
for $\mbox{Im} \, \omega = 0$.   In this section, in addition to ruling out quasinormal modes with real $\omega \neq 0$, we study some other properties of the Green's functions for real $\omega$.

Consider the quantity $W=A' \bar A - \bar A' A$ introduced above, which for real $\omega$ can be associated with the Wronskian. 
As discussed below (\ref{intparts}), a necessary condition for $A$ to be
a quasinormal mode is that $r^n F W$ must vanish in the limit $r \to \infty$.  Recall that the Wronskian satisfies the differential equation
\be
W'(r) + T_1(r) W(r) = 0\, ,
\ee
which for $T_1=(F'/F+n/r)$ is solved by $W\propto r^{-n}/F(r)$ up to an $r$-independent constant.
Assuming 
$A \approx (r-r_h)^{-i \omega / F_h}$ near the horizon, we find
\be \label{eq:horizonW}
r^n F (A' \bar A - \bar A' A) = -i \omega r_h^n\ .
\ee
Thus the Wronskian never vanishes unless $\omega =0$ (or in the limit $r_h\to 0$). We conclude there are no quasinormal modes on the real axis away from the origin for non-zero $r_h$, and hence no poles in $G_R(\omega, k)$ for 
real $\omega \neq 0$.  

To understand the positivity properties of the spectral function, we use the Wronskian to show $\omega \, \rho(\omega, k) \geq 0$ (for real $\omega$).  
The idea is very simple.  Up to an overall normalization,
we can write the spectral density as \cite{Gubser:2008wz}
\be
\mbox{Im} \, G_R(\omega) \sim \frac{-1}{2i} \lim_{r \to \infty} F r^n \left( \frac{A'}{A} - \frac{\bar A'}{\bar A} \right) \ .
\ee
As $A$ and $\bar A$ are linearly independent solutions, this expression is proportional to the Wronskian $W$ for the differential equation (\ref{eq:order2}).
Thus the spectral density can be written in a way that makes its positivity properties manifest:
\be\label{eq:bosonicOne}
\mbox{Im} \, G_R(\omega) \sim \frac{\omega}{2} \lim_{r\to \infty} \frac{r_h^n}{|A|^2} \ ,
\ee
using equation \eqref{eq:horizonW}.  

To understand the reality property $G_R(\omega,k)^* = G_R(-\omega,-k)$ (for real $\omega$), note first that (\ref{eq:order2}) is a real differential equation that depends on $\omega^2$ and $k^2$, not on $\omega$ or $k$ by themselves.  The solution $A$ is complex only because of the near horizon boundary condition (\ref{nearhorizonA}).  Thus $A$, up to an overall constant phase factor, is a function of $i \omega$ with real coefficients.  $A$ is also an even function of $k$.  Given the recipe for constructing $G_R(\omega,k)$ from $A$, the result follows.  

Note there exists a physical interpretation of the quantity (\ref{eq:horizonW}).\footnote{%
Ref.\ \cite{Gubser:2008wz} was the first to notice the importance of this quantity.
}   The action (\ref{SAaction}) contains a conserved Noether charge $Q$ associated with phase rotations of $A$ and $\bar A$.  The left hand side of (\ref{eq:horizonW}) is equal to this $Q$.  Thus ingoing boundary conditions at the horizon force $Q \neq 0$ which in turn guarantees that there are no quasinormal modes for real $\omega \neq 0$.  
In the case where $A$ is a scalar field, 
this charge $Q$ is related to the radial component of a charge current $J^r$.  
A slightly more elaborate argument relates
(\ref{eq:horizonW}) to current conservation $\nabla_a J^a = 0$: $J^t$ is independent of $t$ which in turn forces $\nabla_r J^r = \frac{1}{\sqrt{-g}} \partial_r \sqrt{-g} \,  J^r = 0$.  The left hand side of (\ref{eq:horizonW}) is $\sqrt{-g}  J^r$ up to normalization.  
The sign of
$\mbox{Im} \, G_R(\omega)$ is related to the sign of the charge current $J^r$. 

The absence of poles for real $\omega$ may seem surprising in view of some results in the literature.  
For example, there are Dp/Dq brane systems in black hole backgrounds
that support {\it normal} modes, i.e.\ poles at real $\omega$ at nonzero 
temperature~\cite{Hoyos:2006gb,Kruczenski:2003be}. 
Despite the nonzero temperature,
these systems still have $r_h=0$ in \eqref{eq:horizonW}.\footnote{
In the D3/D7 brane system with $d=4$ at {\it vanishing charge density}
and at sufficiently low temperature,
there are (scalar and vector) fluctuations with modes having real frequencies.
At low $T\neq 0$ the brane ends outside the black hole as opposed to falling
into it. Thus the metric induced on the brane has $r_h=0$ ({\it Minkowski embeddings})
\cite{Babington:2003vm,Kirsch:2004km,Mateos:2006nu}.
}
Another distinct class of examples, where $r_h=0$ coincides with $T=0$, 
has normal modes only in the limit $T \to 0$.  At any 
$T \neq 0$, the modes have at least a small imaginary part.  
The current-current correlation function for the holographic superconductor with a charged scalar at the Breitenlohner-Freedman bound 
is in this class \cite{Horowitz:2008bn}.
Such modes were also observed in 
the low temperature limit of current-current correlation functions for the D3/D7 system, at {\it nonzero
charge densities} \cite{Kaminski:2009ce,Erdmenger:2007ja, Myers:2008cj,Kaminski:2008ai}.  

 Even if we were to allow such a loophole of poles at real $\omega\neq 0$ here, 
 given positivity of the spectral function
$\omega \, \rho(\omega,k) > 0$, provided the poles are single poles, they will not spoil the sum rule.  
The single poles at $\omega \neq 0$ must have purely real residues.\footnote{%
 If the residue at $\omega=c$ has a nonzero imaginary part $b$, then
 $\rho \sim b / (\omega - c)$.  The quantity $\omega \rho$ will change sign as
 $\omega$ passes through the pole. 
}
Regularizing with an $i \epsilon$ prescription, these poles in the real part of $G_R(\omega, k)$ introduce Dirac delta functions into the spectral density which we can integrate over.

\subsection{Contraction Map for Bosonic Correlators}
\label{sec:contractbosons}

To study the large $\omega$ asymptotics of the Green's functions from gravity, we find it convenient first to convert (\ref{eq:order2}) to an equivalent, nonlinear first order differential equation.
If we have functions $P(r), Q(r)$, then we can consider the quantity
\begin{equation}
s \equiv \frac{Q' A - Q A'}{P' A - P A'}.
\end{equation}
(Usually we want $P$ and $Q$ to be approximations of solutions of (\ref{eq:order2}).)
Then
\begin{equation}
s' = -\frac{(sP-Q)\left[s {\mathcal D}(P) - {\mathcal D}(Q) \right]}{PQ'-P'Q}.
\ee
Also, we have the identity
\begin{equation} \label{eq:stosigma}
\frac{A'}{A} = \frac{P's-Q'}{Ps-Q}.
\end{equation}

We define $u=\int_r^{\infty} \frac{dr'}{F}$, assuming that 
\be \label{eq:fexpansion}
F \sim r^2 \left( 1 - p r^{-d} + o(r^{-d}) \right) \ 
\ee
at large $r$
and $F \sim (r-r_h)$ for $r \approx r_h$.\footnote{%
Equation \eqref{eq:fexpansion} does not hold for all AdS theories.  In general it will only
hold if there are no fields with scaling dimension $d/2$ or lower.
}
Having set $L=2 \kappa^2 = 1$, $p$ here can be related to the on-shell value of the action and hence to the pressure of the field theory.  For a review of this claim, see Appendix \ref{app:TT}.
We choose
\begin{eqnarray}
P &=& \frac{u^{1/2}}{r^{n/2}} H_{(n+1)/2}^{(2)}(\omega u) \ , \\
Q &=& \frac{u^{1/2}}{r^{n/2}} H_{(n+1)/2}^{(1)}(\omega u) \ . 
\end{eqnarray}
Note that\footnote{Equating the r.h.s. of equation \eqref{eq:Y} to zero 
determines those $Y$ for which $P$ and $Q$
are {\it exact} solutions. This observation may be useful 
for  constructing
potentials with exact solutions.}
\be\label{eq:Y}
\frac{{\mathcal D}(P)}{P} = \frac{{\mathcal D}(Q)}{Q}
= \frac{n}{2} \left( \frac{2-n}{2 r^2} + \frac{n+2}{2 u^2 F^2} - \frac{F'}{r F} \right) - \frac{Y}{F} \ .
\ee
and that the Wronskian has the form
\be
P Q' - P' Q = -\frac{4i}{ \pi r^n F} \ .
\ee
Thus the equation for $s$ becomes
\be
s' = {\mathcal E}(s) \equiv \frac{\pi}{4iF} \left( s  H_{(n+1)/2}^{(2)} (\omega u) -  H_{(n+1)/2}^{(1)} (\omega u) \right)^2 y
\label{eq:s}
\ee
where we have defined
\be
y \equiv \left( \frac{n}{2} \left( \frac{2-n}{2 r^2} + \frac{n+2}{2 u^2 F^2} - \frac{F'}{r F} \right) - \frac{Y}{F} \right) u F^2 \ .
\ee
We will call the expression multiplied by $n/2$ above the pressure term because 
it depends on $p$ through $F$.  

We 
will now derive an asymptotic expansion for $s(r, \omega)$ in the limit $\omega \to \infty$, for 
$\mbox{Im} \, \omega \geq 0$.
We introduce the mapping 
\be
\mathcal{I}(s) = \int_{r_H}^r \mathcal{E}(s(r')) dr' \ .
\label{contractmap}
\ee
We will assume that  $\mbox{Im} \, \omega \geq 0$ and that 
$y(u) = O(u^{2\Delta - 1})$ as $u \to 0$.  For the current-current correlator, we will assume
$\Delta > (d-2)/2$ (the unitarity bound for scalar operators).  For the stress-tensor correlator, we will assume that $2 \Delta > d$.  (This latter bound forces us to consider sum rules for the stress tensor with $k=0$.)  Given these assumptions, 
we will show in Appendix \ref{app:contractbosons}
that this map (\ref{contractmap}) 
is a contraction mapping with contraction factor $O(|\omega|^{-1})$.  
Therefore, the error in $s$ decreases by a factor
of $|\omega|^{-1}$ each time we iterate $\mathcal{I}$.

Generally, $\Delta$ is the scaling dimension of
a field $\Phi$ that is coupled to $A$. (Because of Lorentz and gauge symmetry constraints, 
both gauge and gravity fields typically couple
to quadratic functions of fields, hence the factor of $2$.)
We are particularly interested in the effect of a nonzero 
pressure $p$, for which the effective 
$\Delta = d/2$.  
(In the stress tensor case $n=d-1$, the leading contribution from the
pressure vanishes and $\Delta > d/2$.)
Although our emphasis on the gravity side of the duality obscures the point, one should be able to think of these couplings in terms of an operator product expansion (OPE) 
on the field theory side.  
The effect of $p$ on the large frequency behavior of the correlation function comes from the presence of $\langle T^{\mu\nu} \rangle$ in an OPE of the current-current or stress tensor-stress tensor correlation functions.  Similarly, the effect of a quadratic coupling to a field $\Phi$ of dimension $\Delta$ corresponds to the presence of $\langle {\mathcal O}{\mathcal O} \rangle$ in the OPE 
(where $\Phi$ is dual to ${\mathcal O}$).

\subsection{Explicit Computation of Correlators}

Now, we actually compute the asymptotics of the correlators in question.

The first observation is that the large $\omega$ behavior of $s$ is dominated by the large $r$ region of the differential equation.  This observation is intuitive from the AdS/CFT correspondence because large $\omega$ behavior corresponds to the UV of the field theory, and the UV of the field theory is precisely this large $r$ region.  From a mathematical point of view, 
we notice that $H_{(n+1)/2}^{i}(\omega u)$ oscillates very rapidly
when $\omega$ is large.
If we expand a rapidly oscillating integral
$\int_0^{\infty} f(u) g(\omega u) du$
as an asymptotic series in powers of $\omega$, and $f$ and $g$ are $C^{\infty}$ functions,
then only their behavior near $0$ (the AdS boundary) and $\infty$ (the black hole horizon) 
is important.
(One way to see this is to integrate by parts repeatedly.)
In our case, $f=\frac{n+2}{u^2}$ plus some terms that are multiplied
by $F$ and therefore decrease exponentially for large $u$.
So all derivatives of $f$ go to zero at the horizon. 

In the proof of the contraction mapping above, we neglected suppression due to these oscillations.  Typically, we expect that $s$ will receive contributions only from the large $r$ region of the integral and so be of order $\omega^{-2\Delta}$.  If the pressure term dominates, then $2 \Delta = d$ while if the $Y$ term dominates, then $\Delta$ will typically be the dimension of a scalar operator contributing to $Y$, for example the $\Phi$ in (\ref{Axeq}).

We will consider the current correlators in $d=3$ and $d=4$ and then the stress tensor correlator in general $d$.  We begin with the current two point function in $d=3$.  From combining (\ref{GRxxdef}) and (\ref{eq:stosigma}), we have that
\be\label{eq:GRxx}
G_R^{xx}(\omega,0) =
i  \omega \frac{1-s}{1+s} \ ,
\ee
(setting $e=1$).
Note that when $d=3$, then $n=0$ and there is no contribution to $s$ from the pressure term in (\ref{eq:s}).  
We have from the preceding discussion that a nonzero $Y$ will induce an $s \sim \omega^{-2\Delta}$.  
We expect that the dimension should exceed the unitarity bound $2\Delta > 1$.  
Thus, we expect the following variant of the Ferrell-Glover-Tinkham sum rule to hold for a generic choice of $Y$:
\be
G_R^{xx}(0,0) = \frac{2}{\pi} \int_0^\infty \frac{d\omega}{\omega} \left( \mbox{Im} \, G_R^{xx}(\omega,0) -  \omega  \right) \ .
\ee 

For the current two point function in $d=4$, we find that
\be
G_R^{xx}(\omega,0) = 
\omega^2
\left( \ln (i \Lambda/ \omega) - \frac{i \pi s}{1+s} \right) \ .
\ee
The contribution to $s$ from the pressure term should scale as $1/\omega^4$ while that from $Y$ should scale as $\omega^{-2\Delta}$ where now from unitarity $\Delta > 1$.  Thus a 
Ferrell-Glover-Tinkham sum rule of the form 
\be
G_R^{xx}(0,0) = \frac{2}{\pi} \int_0^\infty \frac{d\omega}{\omega} \left( \mbox{Im} \, G_R^{xx}(\omega,0) - \frac{\pi \omega^2}{2} \right) \ 
\ee
should hold for generic choice of $Y$.

We can determine the leading contribution to $s$ from the pressure term for this $d=4$ case.
If $F = r^2 - p r^{2-d} + ...$ for some constant $p$,
then we have $\frac{3}{4Fu^2}+\frac{F}{4r^2} - \frac{F'}{2r} = (-4p/5) u^4 + ...$.  Then we can integrate
\begin{equation}
\int_{0}^{\infty} \frac{-4p u^4}{5} \frac{-i \pi u}{4} H_{1}^{(1)} (\omega u)^2 \frac{du}{u^2} = \frac{8i}{15\pi} p\omega^{-4}.
\end{equation}
(The integral diverges when $\omega$ is real, but we may analytically
continue from the upper half plane, where it does converge.)
So the leading correction to 
$G_R^{xx}(\omega,0)$
from the pressure term is
$(-i\pi \omega^{2})\left(\frac{8i}{15\pi}p \omega^{-4}\right) =
(8p/15) \omega^{-2}$. 
This term agrees with an analytically solvable case discussed
in Appendix \ref{app:d4}.  See in particular (\ref{largeomegad4JxJx}).
By dimensional analysis, this correction is of the form one would expect if the first nontrivial term in the OPE of $J^x$ with itself contains the stress-tensor.

For the stress tensor correlator, we are interested in the component $G_R^{xy,xy}(\omega,0) = G_R(\omega)$. 
Combining the results of Appendix \ref{app:TT} and (\ref{eq:stosigma}), we find that 
\be
G_R(\omega) = -p - \frac{3p}{4} + i \omega^3 -  \frac{2 i \omega^3 s}{1+s} \ ,
\ee
\be
G_R(\omega) = -p - \frac{6p}{5} + \frac{1}{4} \omega^4 \ln (i \Lambda / \omega) - \frac{i \pi s \omega^4 }{4 (1+s)} \ ,
\ee
in $d=3$ and $d=4$ respectively.    Similarly to the current-current correlation function, the presence of $p$ in the expressions above is consistent with what one would expect if the OPE of the stress-tensor with itself contains the stress-tensor.
To eliminate the power law growth with respect to $\omega$ and construct a sum rule, we consider the regulated Green's function
\be
\delta G_R(\omega) \equiv G_R(\omega,T) - G_R(\omega, 0) \ ,
\ee
where $p$ and $s$ will generically depend on temperature $T$.
The leading $p$ in $G_R(\omega)$ comes from a contact term that we determined in Appendix \ref{app:TT}.  The second, $p$ dependent term comes from a correction to $Q$.  
This correction can be generalized to arbitrary $d$.  
We have to look at the large $r$ limit of $Q$:
\be
Q = (ur)^{(1-d)/2} u^{d/2} H_{d/2}^{(1)} (\omega u) \ .
\ee
and in particular we have $(ru)^{(1-d)/2} = 1 - \frac{p(d-1)}{2(d+1)} r^{-d} + O(r^{-d-2})$. 
So $\frac{Q - Q_{p=0}}{Q_{p=0}} = \frac{-p(d-1)}{2(d+1)} r^{-d}$ and 
$\delta G_R(\omega) = -p - p (d-1) d/ 2 (d+1) + O(s)$. 

To understand the correction from $s$, note that having set $k=0$ (to be consistent with the bounds on $\Delta$ and have a contraction map), 
we find $Y=0$ in (\ref{eq:s}).  As mentioned above, in the special case $n=d-1$ the leading contribution to $s$ from the pressure term vanishes.  We would need to specify more terms in a power series expansion for $F$ to compute $s$ reliably, but we know that 
$s \sim \omega^{-2 \Delta}$ where $\Delta > d/2$.  Thus $s$ will lead to corrections of 
$G_R(\omega)$ that vanish in the limit $\omega \to \infty$.
We find 
\be
\lim_{\omega \to \infty} \delta G_R(\omega) = -p - \frac{p (d-1) d}{ 2 (d+1)} \ .
\ee
To understand the sum rule, we also need the behavior of $G_R^{xy,xy}(0,0)$.  
When $\omega=0$, the equation (\ref{eq:order2}) is easy to solve, $A = $ const.  
Thus from Appendix \ref{app:TT}, $G_R^{xy,xy}(0,0) = -p$, and
we find the sum rule
\be
\frac{1}{\pi} \int \frac{d \omega}{\omega}  \delta \rho(\omega)= \frac{(d-1)d}{2(d+1)} p \ .
\label{TTsumrule}
\ee
In $d=4$ we reproduce the result of Romatschke and Son
\cite{Romatschke:2009ng}:
$\frac{1}{\pi} \int \frac{d\omega}{\omega} \delta \rho(\omega) = \frac{6}{5}p = \frac{2}{5}\epsilon$.  (In the last equality, we used the fact that for a conformal field theory, the stress-tensor is traceless and so the energy density $\epsilon = 3p$.)  Note we expect this result (\ref{TTsumrule}) to hold in arbitrary $d$ for a generic $F$ satisfying our assumptions.

\section{Fermions}
\label{sec:fermions}
We now repeat our analysis carried out in the previous section for the case of Green's functions of fermionic operators. Again, it has to be shown that the quasinormal modes 
of (now fermionic) bulk fields exclusively lie in the lower half of the complex plane 
and that the spectral function falls off fast enough at large frequencies. 
Our strategy is to decompose the Dirac equation into a set of identical
differential equations involving two-component spinors. This decomposition
allows us to study sum rules in arbitrary dimensions in a simple way.

\subsection{Fermionic Green's functions in arbitrary dimensions}
Consider a fermionic operator ${\mathcal O}_\Psi$ in a field theory dual to a 
spinor $\Psi$ in gravity with the action
\be\label{eq:Sfermion}
S_\Psi = -i \int d^{d+1} x \, \sqrt{-g}\,  \overline \Psi (\gamma^a D_a
- \Phi) \Psi \ ,
\ee
where $\Psi$ is a Dirac spinor of charge $q$, $\overline \Psi = \Psi^\dagger \gamma^{\underline t}$.
$\Phi$ is a real scalar field which allows us to study fermionic fluctuations in presence of 
a scalar condensate, e.g. in a holographic superconductor.\footnote{%
This kind of coupling along with others, and their various implications,
were discussed in~\cite{Faulkner:2009am}.
}
We do not separate $\Phi$ from a possible fermion mass $m$.
When we eventually take the metric to be of the form (\ref{eq:metric0}), 
we will assume that $\Phi$ goes to zero at the 
boundary as $mL +\Phi_0 r^{-\Delta}$ for some $\Delta > 1$ (which is in general different from the 
bound on $\Delta$ in the previous section) and that $\Phi$ is finite at the black hole horizon.
For the sake of generality we keep 
the fermion mass $m \neq 0$ for now, but will set it to zero soon below.
Furthermore, we assume that
$A_t$ is the only nonzero component of $A$.
The covariant derivative is
\be
D_a = \partial_a + \frac{i}{4} \omega_{a, \underline{bc}} \gamma^{\underline{bc}} - i q A_a \ ,
\ee
where $\omega$ is the spin connection and $A_a$ is a gauge field. For a diagonal metric that depends only on the radial coordinate $r$, the spin connection 
can be written
\be
\omega_{\nu, \underline{\mu r}} = \eta_{\mu\nu} \sqrt{g^{rr}} \partial_r \sqrt{|g_{\mu\nu} |} \ .
\ee
The gamma matrices $\gamma^a$ satisfy the Clifford algebra $\{ \gamma^a, \gamma^b \} = 2 g^{ab}$.  Vielbein indices are underlined, and the generators of the Lorentz group are $\sigma^{\underline{ab}} = \frac{i}{2} [ \gamma^{\underline a}, \gamma^{\underline b} ] = i\gamma^{\underline{ab}}$.  The dimension of ${\mathcal O}_\Psi$ is $d/2\pm mL$,
where $\Delta_\Psi \ge (d-1)/2$ satisfies the unitarity bound \cite{Minwalla:1997ka}.

\paragraph{Dirac equation in arbitrary dimensions}
Assuming the metric is diagonal and depends only on the radial coordinate $r$, it is convenient to rewrite the equations of motion in terms of a rescaled $(2^n)$-component spinor $\psi = (-g\cdot g^{rr})^{1/4} \Psi$ where $n = \lfloor (d+1)/2 \rfloor$.  Given the above assumptions on the form of the metric, the equation of motion for $\psi$ is
\be
[\gamma^a (i \partial_a + A_a)-i \,
\Phi
] \psi = 0 \ ,
\ee 
setting $q=1$.

We suppose that the fermion has some momentum $k$.  Because of rotational symmetry, we may assume
without loss of generality that the momentum is in the $x$ direction.
The equation of motion for $\psi$ is then
\be \label{eq:fermion}
\left[ \sqrt{-g^{tt}} \gamma^{\ut} (\omega + A_t) +i \sqrt{g^{rr}} \gamma^{\ur} \partial_r -
\sqrt{g^{xx}} \gamma^{\ux} k -i \, 
\Phi
\right] \psi = 0.
\ee 

We argue that (\ref{eq:fermion}) can be decomposed into a set of identical differential equations acting on two-component spinors.  Consider a Euclidean Clifford algebra, $\{ \tilde \gamma^{a}, \tilde \gamma^{b} \} = \delta^{ab}$ in $D$ dimensions.  In the case $D=2$, we take $\tilde \gamma^0 = \sigma_3$ and $\tilde \gamma^1 = \sigma_2$ the Pauli spin matrices.  Given a Clifford algebra $\tilde \gamma^a$ in $D=2n$ dimensions, a Clifford algebra $\tilde \Gamma^a$ in $D+2=2n+2$ dimensions is
\be
\tilde \Gamma^{0} = \mbox{id} \otimes \sigma_3 \ , \; \; \;
\tilde \Gamma^{1} = \mbox{id} \otimes \sigma_2 \ , \; \; \;
\tilde \Gamma^a = \tilde \gamma^{a-2} \otimes \sigma_1 \ .
\ee
For odd dimensions, in the usual way we identify $\tilde \gamma^{2n+1}$ with the product of the other gamma matrices (up to a factor of $i$). Given this construction, we can think of $\tilde \gamma^0$ and $\tilde \gamma^{1}$
as block diagonal, consisting of $\sigma_3$ or $\sigma_2$ matrices along the diagonal.  Additionally,
$\tilde \gamma^{2}$ (for $D>2$) is block diagonal, consisting of $\sigma_1$ matrices along the diagonal of alternating sign.
The Lorentzian Clifford algebra can be recovered by multiplying one of the $\tilde \gamma^a$ by $i$.  We choose $\gamma^{\ut} = i \tilde \gamma^{1}$, $\gamma^{\ur} = \tilde \gamma^0$ and $\gamma^{\ux} = \tilde \gamma^{2}$.  With these choices, (\ref{eq:fermion}) in an arbitrary number of dimensions $d$ reduces to two decoupled equations
\be
\label{Diraceqsimp}
\left[ \sqrt{-g^{tt}} i \sigma_2 (\omega + A_t) +i \sqrt{g^{rr}}\sigma_3 \partial_r +(-1)^\alpha
\sqrt{g^{xx}} \sigma_1 k -i \, 
\Phi
 \right] \psi_\alpha = 0\, ,
\ee
where $\psi_\alpha$ with $\alpha=1,2, \dots\, , 2^{n-1}$ 
are the $2^{(n-1)}$ two-component spinors appearing in the 
$2^{(n-1)}$ spinor equations, respectively, with momentum alternating between $+k$ and $-k$. 
Therefore the second block of  equations, i.e. the one for $\psi_2$, is related to the first block by $k\to -k$.  A third block (if present) is identical to the first, the fourth block to the second, and so on.

\paragraph{Fermion operator Green's functions}
We are interested in the retarded Green's function for the boundary
fermion operator $\mathcal{O}_{\Psi}$.
The prescription we use to study this fermion correlation function closely follows that of \cite{Iqbal:2009fd,Liu:2009dm,Faulkner:2009wj,Cubrovic:2009ye} which is based on the
work of \cite{Son:2002sd}.  In what follows we will summarize this prescription.
Consider a solution to the Dirac equation with infalling boundary conditions at the black hole horizon. 
Before setting $mL=0$ below, let us consider small masses $mL<1/2$ now.
Decomposing $\psi$ into eigenvectors of $\gamma^r$,
for $mL<1/2$ we find the AdS-boundary behavior  
\be \label{eq:boundaryExp}
\psi_\alpha = 
\left( \begin{array}{c}
\psi_{\alpha, +}\\
\psi_{\alpha, -}
\end{array}\right) =
\left( \begin{array}{c}
a_{\alpha} \, r^{+mL} + {\mathcal{O}}(r^{mL-1},r^{-mL-1})\\
b_{\alpha} \, r^{-mL} + {\mathcal{O}}(r^{-mL-1},r^{mL-1})
\end{array}\right)\, , \quad \alpha = 1,\, 2, \,\dots,\, 2^{(n-1)} .
\ee
We are going to work in the {\it massless} limit from now on, so $mL=0$
and the two spinor components scale with the same power of $r$ near the 
boundary.\footnote{%
 Note that the boundary expansion \eqref{eq:boundaryExp} can involve logarithmic
 sources for half integer masses.
 Furthermore, for $mL >1/2$ the term involving $b_\alpha r^{-mL}$ would become subleading
 compared to $\mathcal{O}(r^{mL-1})$. 
} 

Note that for $mL=0$ two quantizations are possible, choosing either $a_\alpha$
or $b_\alpha$ as the sources.
Here we choose to identify the number $a_\alpha$ with the source for the operator ${\mathcal O_\alpha}_\Psi$, while $b_\alpha = \langle {\mathcal O}_{\alpha \Psi} \rangle$. Since the Dirac equation is linear, $a_\alpha{(\omega,\vec k)}$ will be linearly related to $b_\beta{(\omega,\vec k)}$.
In order to make this relation explicit, we define the $2^{n-1}$ component spinors
$\psi_\pm$.  At the boundary,  $\psi_+$ asymptotes to the sources, $\psi_-$ to the vevs:
\be
\lim_{r\to \infty} \psi_+ = a \; ; \; \; \;
\lim_{r \to \infty} \psi_- = b \ .
\ee
The linear relation between vevs and sources can be written
\be
b{(\omega,\vec k)} = {\mathcal S}\ a{(\omega, \vec k)}  \ .
\ee
The implicit prescription to compute the retarded Green's function is then
\cite{Iqbal:2009fd}:
\be
b = - i G_R \gamma^t_{\rm bdy} M a \ .
\label{eq:implicitGR}
\ee
Note that $\gamma^t_{\rm bdy}$ is the field theory gamma matrix.  $M$ is a change of basis matrix that allows for $a$ and $b$ to transform in different representations of the Clifford algebra.
In odd dimensional AdS spaces, $\gamma^t_{\rm bdy} = \gamma^{\ut}|_+$, i.e. $\gamma^{\ut}$ restricted to the positive eigenspace of $\gamma^{\ur}$,  and $M=\rm{id}$.  
In this case $G_R$ is defined only on the negative eigenspace of $\gamma^{\ur}$ 
and $\gamma^{\ur}$ itself is reinterpreted as 
the boundary gamma matrix which determines the chirality of the field theory spinors.  With our boundary conditions, the expectation values $b$ are negative chirality Weyl spinors.  Choosing the other quantization would allow us to study the opposite chirality.

For even dimensional AdS spaces where $d+1=2n$, we need to construct the field theory gamma matrices.  One simple choice is to let $b_\alpha$ transform under
\begin{eqnarray*}
\gamma^t_{\rm bdy} &=& i^{n+1} \tilde \gamma^0 \cdots \tilde \gamma^{d} \ , \\
\gamma^c_{\rm bdy} &=& \tilde \gamma^c \gamma^t_{\rm bdy} \ .
\end{eqnarray*}
such that $a_\alpha$ is related to $b_\alpha$ via $M = \gamma^t_{\rm bdy}$.  With this choice, the Lorentz generators $\gamma^{\lambda\nu}_{\rm bdy}$ and $\gamma^{\ulambda \unu}$ are compatible.

We rewrite the correlator recipe \eqref{eq:implicitGR}, 
\be\label{eq:fermionGR}
[G_R(\omega,\vec k)]\du{\alpha}{\alpha}= -i \frac{b_\alpha}{a_\alpha}\, ,\quad \alpha \in \{1,\,2,\,\dots,\, 2^{(n-1)}\} \;.
\ee
Note that in the present setup the matrix-valued Green's function is diagonal, 
because the spinors $\psi_\alpha$ decouple due to the block diagonal structure
of the Dirac equation discussed above. We now analyse the pole structure of 
the fermionic Green's functions from the gravity point of view in analogy to
the bosonic case in the previous section.

\subsection{$\mathrm{Im}\, \omega >0$}
\label{sec:omegaGreater0Fermions}
Analogous to the bosonic case discussed in Section \ref{sec:omegaGreater0} we can again argue
for the absence of branch cuts as shown in Appendix \ref{app:branchCutsFermions}.
Thus the only non-analyticities to consider are again the poles of $G_R$ which now correspond to quasinormal modes of $\psi$.
As was the case with the bosons, there are no poles in the upper half of the complex
plane ($\mbox{Im}\, \omega >0$).  Quasinormal modes are solutions to (\ref{eq:fermion})
that are normalizable at the boundary ($a_\alpha=0$) and ingoing at the horizon.  
In the upper half plane, the solutions are also
normalizable at the horizon.  So quasinormal modes in the upper half plane are
eigenfunctions of the Dirac operator of (\ref{eq:fermion}) with eigenvalue $\omega$.
But in the upper half plane the Dirac operator is self-adjoint and can have only real eigenvalues.

As mentioned in Section \ref{sec:holoTheories}, because $\Psi$ is not Hermitian, the retarded Green's function is not expected to have a reality property of the form $G_R^{\alpha \beta} (\omega,k)^* = G_R^{\alpha \beta} (-\omega,-k)$.  For our fermion, however, the Dirac equation (\ref{Diraceqsimp}) has a symmetry under complex conjugation plus $\omega \to -\omega$, $k \to -k$ and $A_t \to -A_t$.  Thus the diagonal components of the Green's functions are expected to have the symmetry 
$G_R(\omega, k, \mu)^* = G_R (-\omega,-k,-\mu)$ where $\mu$ is the boundary value of $A_t$.

\subsection{Real $\omega$}
As was true for the bosonic Green's functions, we now show that there are
no poles on the real axis, {i.\,e.\,}no (quasi)normal modes with real frequencies. 
The fermion spectral function can be written in terms of bulk field components
\be
\rho(\omega, k)  \du{\alpha}{\alpha} = \frac{1}{2 i} \left[{G_R} \du{\alpha}{\alpha} - {({G_R}\du{\alpha}{\alpha})}^\dagger\right]
 = -\lim_{r \to \infty} \frac{1}{2 }\left[ \frac{\psi_{\alpha,-}}{\psi_{\alpha,+}}  + \frac{{\psi_{\alpha,-}}^*}{{\psi_{\alpha,+}}^*} \right]
\, , \quad \, \omega, k \in \mathbf{R} \, .
\ee

The spinor components $\psi_{\alpha,\pm}$ are chosen such that
near the horizon they behave as
$\psi_{\alpha,\pm}\approx (r-r_h)^{-i\omega/F_h}[c_{\alpha,\pm} + \mathcal{O}(r-r_h)]$. 
 $\rho$ has poles if ${\psi_{\alpha,+}}$ and $\psi_{\alpha,+}^*$ vanish simultaneously. 
 The quantity $W_f = (\psi_{\alpha,-}\psi_{\alpha,+}^* + \psi_{\alpha,+}\psi_{\alpha,-}^*)$
is the fermionic analog of the Wronskian considered in Section~\ref{sec:gravity}, 
and has to vanish at a pole. 
It follows from the equations of motion that $W_f$ is independent of the radial coordinate, $\partial_r W_f = 0$ 
(see {e.\,g.\,}~\cite{Kaminski:2009dh, Hartman:2010fk,Gubser:2010dm}).\footnote{%
Similar to the bosonic case, 
this quantity can be related to the radial component of a conserved current
$J^r = \overline \Psi \gamma^r \Psi = W_f / \sqrt{-g}$.  That $\partial_r W_f = 0$ follows from the fact that $\nabla_a J^a = 0$ and the observation that the components of $J^\mu$ depend only on $r$.
\label{fermionJ}
}
Exactly at the horizon we find for the diagonal entries
\be\label{eq:horizonWf}
\lim_{r\to r_h} W_f = -2\, |c_{\alpha,+}|^2 = -2\, |c_{\alpha,-}|^2\, ,
\ee 
because, at $\omega\not =0$, the leading coefficients are fixed
by the infalling boundary condition to satisfy $c_{\alpha,-}=-c_{\alpha,+}$. Thus there are never poles in the fermionic Green's function along the real axis away from $\omega=0$
($c_{\alpha,\pm}=0$ is the trivial case $\psi_{\alpha,\pm}(r)\equiv 0$). 
In contrast, at $\omega=0$ the infalling boundary condition and equations of motion no longer imply a relation between the leading coefficients $c_{\alpha,\pm}$. 
In this case, without loss of generality one of these coefficients can be chosen to be real, the other
imaginary, and
the sum $W_f=[c_{\alpha,-}^* c_{\alpha,+} + c_{\alpha,-} c_{\alpha,+}^*]$ vanishes.
For example, in the zero temperature limit a fermionic (quasi)particle pole
sits at $\omega=0$ at the Fermi momentum $k=k_F$ \cite{Liu:2009dm}.

Positivity of the spectral function can be made manifest using equation 
\eqref{eq:horizonWf} \cite{Gubser:2010dm}. Consider that $\rho\du{\alpha}{\alpha} =-\lim_{r\to\infty} W_f/(2 |\psi_{\alpha,+}|^2)$,
which implies for the diagonal elements that
\be \label{eq:positivityRhoF}
\rho\du{\alpha}{\alpha} = \lim_{r\to \infty} \frac{|c_{\alpha,+}|^2}{|\psi_{\alpha,+}(r)|^2} \ge  0\, .
\ee
Note, that the fermion spectral function \eqref{eq:positivityRhoF} does not 
contain any explicit factor of $\omega$, in contrast to the bosonic one \eqref{eq:bosonicOne}. Therefore,
it is non-negative also for negative frequencies.\footnote{%
 Similar to the bosonic case, the sign of the spectral density is related to the direction of 
 the radial current $J^r$ mentioned in footnote \ref{fermionJ}.
}

\subsection{Asymptotic Series for the Correlator}

Again, we want to determine the large frequency behavior of the Green's 
function for $\mbox{Im} \, \omega \geq 0$. 
In order for a sum rule to be valid, we need Green's functions that
fall off fast enough at large frequencies. In order to examine this large frequency
behavior, we are going to introduce a map $\mathcal{I}(\psi)$.
Let us first define two helpful abbreviations
\be
u = -\int_r^{\infty} dr'  \sqrt{-g^{tt} g_{rr}}\, ,
\ee
and
\begin{equation}
v = -\int_r^{\infty} dr' \sqrt{-g^{tt} g_{rr}}\, A_t.
\end{equation}

While the method that we used for bosons can also be applied to fermions,\footnote{%
 Suppose $A, P, Q$ are functions that take values in a two-dimensional vector space,
 and $A$ satisfies $\mathcal{D}(A)\equiv A'+\mathfrak{T}A = 0$.  
 We define $s=\frac{Q \wedge A}{P \wedge A}$.
 Then $s' = \frac{1}{P \wedge Q} \left[ (P \wedge \mathcal{D}(P))s^2 -
 (P \wedge \mathcal{D}(Q) + Q \wedge \mathcal{D}(P))s + (Q \wedge \mathcal{D}(Q)) \right]$.
 The second order system $A''+{T}_1 A'+{T}_0 A=0$ is just a special case
 with $A \to \twobyone{A}{A'}$, $P \to \twobyone{P}{P'}$, $Q \to \twobyone{Q}{Q'}$, and
 $\mathfrak{T} \to \twobytwo{0}{-1}{{T}_0}{{T}_1}$.
} 
we choose to use a different method that can be generalized
more readily to spinors with more than two components.
We define the mapping
\begin{equation}\label{eq:I}
\mathcal{I}(\psi) = e^{-i (\omega u+v)} \tilde{\psi}^{(0)} +\mathcal{J}(\psi)\, ,
\end{equation}
with
\begin{equation}
\mathcal{J}(\psi)= i e^{i \gamma^{\ur} \gamma^{\ut} (\omega u + v(u))} \int_{u}^{\infty} du' e^{-i\gamma^{\ur} \gamma^{\ut} (\omega u'+v)} \gamma^{\ur} \left(\sqrt{\frac{g^{xx}}{-g^{tt}}}k \gamma^{\ux} + \frac{i \Phi}{\sqrt{-g^{tt}}} \right) \psi
\end{equation}
where $\tilde{\psi}^{(0)}$ satisfies
\begin{equation} \label{eq:fermioncontraction}
\gamma^{\ur} \gamma^{\ut} \tilde{\psi}^{(0)} = -\tilde \psi^{(0)},\qquad \partial_u \tilde{\psi}^{(0)} = 0.
\end{equation} 
Then a fixed point of $\mathcal{I}$ is a solution to (\ref{eq:fermion}) with
infalling boundary conditions. In other words $\mathcal{I}(\psi) = \psi$ is equivalent
to the equation of motion \eqref{eq:fermion}, as can be shown by taking the
derivative with respect to $u$ on both sides of $\mathcal{I}(\psi) = \psi$.

We can write the first term of \eqref{eq:I} as
\be\label{eq:exactSolution}
\psi^{(0)}=\mathcal{I}(0) = e^{-i (\omega u+v)} \tilde{\psi}^{(0)}\, .
\ee
This is a solution to the equation of motion \eqref{eq:fermion} for 
$k=0$ and $\Phi\equiv 0$. 
Therefore it can also be considered an 
approximate solution to the full equation \eqref{eq:fermion}. 
We will show in Appendix \ref{app:contractfermions} that the map ${\mathcal I}(\psi)$ can be iterated, using
$\psi^{(0)}$ as an initial approximate solution, to find the exact solution.   

Now we actually compute the first correction to the fermion correlator in the large $\omega$ limit.  
The leading term comes from
a ratio constructed from the components of $\psi^{(0)}$ and 
$\psi^{(1)} \equiv {\mathcal I}(\psi^{(0)}) - \psi^{(0)}$. 
We compute $\psi^{(1)}$ and obtain
\be
\psi^{(1)} = i e^{i (\omega u + v)} \int_{u}^{\infty} du' e^{-2i(\omega u'+v)} \gamma^{\ur} \left(\sqrt{\frac{g^{xx}}{-g^{tt}}}k \gamma^{\ux} + \frac{i\Phi}{\sqrt{-g^{tt}}} \right) \tilde{\psi}^{(0)}. 
\label{psi1}
\ee

Once again we have an integral that oscillates rapidly for large $\omega$.
As long as all derivatives of
$\frac{e^{-2iv}}{\sqrt{-g^{tt}}}\gamma^{\ur} \left(\sqrt{g^{xx}} k\gamma^{\ux}+i\Phi \right)$
go to zero at the horizon, the asymptotic series depends only on the behavior near the
boundary.  We have $\sqrt{-g^{tt}} \sim \sqrt{g^{xx}} \sim u$.
If we let $A_t \sim \mu$, then $v \sim \mu u$, so $v \rightarrow 0$ at the boundary.
We also let $\Phi \sim \Phi_0 u^{\Delta}$ with $\Delta>1$.

The leading correction from $k$ is
\be
\psi^{(1)}_k|_{u=0} \sim i \frac{k}{2i\omega} \gamma^{\ur} \gamma^{\ux} \tilde{\psi}^{(0)}.
\ee
The leading correction from $\Phi$ is
\be
\psi^{(1)}_\Phi|_{u=0} \sim  i \frac{\Phi_0}{(2i \omega)^{\Delta}} \Gamma(\Delta) \gamma^{\ur} \tilde{\psi}^{(0)}.
\ee
In our gamma matrix basis, we choose $\tilde \psi^{(0)} = (1,1,\ldots, 1) \otimes (-1,1)$.  
We find then that $\psi^{(1)}|_{u=0} \sim (k/2\omega)(1,-1,1,-1,\ldots,1,-1)\otimes(1,1) + o(1/\omega)$.
Therefore at large $\omega$
\be
{G_R(\omega,k)^\alpha}_\alpha = \left( i \pm \frac{i k}{\omega} + o(1/\omega) \right) \ .
\ee
Note that if we consider instead the trace (provided $d>2$), 
\be
\mbox{tr} \, G_R(\omega,k) = 2^{n-1} i + o(1/\omega) \ .
\ee
Thus, we can write down a sum rule of the form
\be \label{eq:fermionicSumRule}
\lim_{\omega \to 0} \omega \, \mbox{tr} \, G_R(\omega,k) = \int \frac{d \omega}{\pi } (\mbox{tr} \,\rho(\omega,k) - 2^{n-1} ) \ .
\ee
The left hand side will typically vanish provided $\mbox{tr} \,G_R(\omega,k)$ is finite at the origin. 
This sum rule is similar in spirit to but different in detail from a fermionic sum rule that arises in the context of ARPES (Angle Resolved Photo Emission Spectroscopy) 
\cite{Damascelli:2003bi}. This ARPES sum rule is $\int d \omega A(\omega, \mathbf{k}) = 1$, with the spectral function $A(\omega, \mathbf{k})$.

\section{Discussion}
\label{sec:discussion}

We have seen how some well established properties of Green's function in field theory are realized in a dual gravitational description through the AdS/CFT correspondence.  While our results are not completely general, they do apply to a class of 
correlation functions involving scalars, conserved currents, the stress tensor, and massless charged fermions.
Some of these properties were easy to establish and the results were 
partially known in the literature before.  For example the positivity properties of the spectral function, $\omega \rho (\omega, k) \geq 0$ for bosons and $\rho(\omega, k) \geq 0$ for fermions, follow from the horizon boundary conditions and a conserved charge.
That  $G_R(\omega,k)^* = G_R(-\omega, -k)$ for Hermitian operators turns out to be an almost trivial feature of the differential equation and horizon boundary condition governing the Green's function (and similarly $G_R(\omega,k,\mu)^*=G_R(-\omega,-k,-\mu)$ for fermions).  
Other results were more significant.  For the bosons and fermions, 
we established necessary conditions for 
$G_R(\omega, k)$ to be holomorphic in the upper half of the
 complex frequency plane including real $\omega \neq 0$. 
We developed a contraction map for the bosons and a contraction-like map for the fermions
that allowed us to determine the large frequency behavior of $G_R(\omega,0)$.  We suspect that these contraction maps may be useful in other contexts to study the large frequency behavior of Green's functions from gravity.

We were able to rederive some known current and stress tensor sum rules \cite{Romatschke:2009ng,Hartnoll:2008vx,Baier:2009zy} but from a different viewpoint and in a more general context.  From the results of this paper, we see that these sum rules hold not just for black D3-brane backgrounds \cite{Romatschke:2009ng, Baier:2009zy} and not just for the holographic superconductor in $AdS_4$ with a scalar operator of conformal dimension one or two \cite{Hartnoll:2008vx}, 
but more generally, in other space-time dimensions and in cases where the backgrounds may be deformed by expectation values of arbitrary operators (though typically some lower bound was placed on the conformal dimension of these operators).

One interesting observation that follows from generalizing the sum rule of \cite{Romatschke:2009ng} to arbitrary dimension
is that (\ref{TTsumrule}) for the stress-tensor correlator $G^{xy,xy}(\omega,0)$ can be related to a similar sum rule for 
Chamblin-Reall backgrounds \cite{Springer:2010mf}.  
The sum rule of \cite{Springer:2010mf} can be written in our notation as
\be
\frac{sT}{2} \frac{1}{1 + 2 v_s^2} = \frac{1}{\pi} \int \frac{d \omega}{\omega} \delta \rho(\omega) \ .
\label{TTsumrulealt}
\ee
As discussed by \cite{Gubser:2008ny}, these five dimensional 
Chamblin-Reall backgrounds can be realized as dimensional reductions of asymptotically $AdS_{d+1}$ space-times on a $d-4$ dimensional torus.  The speed of sound $v_s$ and temperature $T$ of the Chamblin-Reall background are  identified with those of the $AdS_{d+1}$ space-time, while the entropy density $s$ and pressure $p$ are rescaled by the volume of the torus.  For conformal field theories dual to these $AdS_{d+1}$ space-times, we expect $sT = p d$ and $v_s^2 = 1/(d-1)$.  Making these substitutions converts (\ref{TTsumrulealt}) into (\ref{TTsumrule}).

As pointed out by \cite{Kanitscheider:2009as}, this trick of dimensional reduction also works for constructing black D$p$ brane solutions from higher dimensional $AdS_{d+1}$ space-times.  We find that black D1 and D4 brane solutions come from reducing $AdS_4$ and $AdS_7$ space-times on a circle, as one might expect given the relation between M-theory and type IIA string theory.  More formally, D0 branes and D2 branes come from $AdS_{d+1}$ where $d=14/5$ and $d=10/3$ respectively.  Thus we find predictions of Romatschke-Son type sum rules for black D0, D1, D2, and D4 branes.  In general, D$p$-branes are related to dimensional reductions of 
$AdS_{d+1}$ space-times with $d = 2(p-7)/ (p-5)$ \cite{Kanitscheider:2009as}.

In the appendices, we wrote down some mildly interesting sum rules that to our knowledge are new.  
We noted that the current correlation function in $d=2$ has a pole at $\omega=0$ that makes the integral in the naive sum rule (\ref{genericsumrule}) divergent.  One possible construction that evades this problem is (\ref{frullani}).  Another more standard sum rule (\ref{d2gensum})  involves regulating the pole by 
taking derivatives.
We also wrote down some new sum rules (\ref{d4gensum}) for the current-current correlation function in the $d=4$ case, i.e. the case of ${\mathcal N}=4$ super Yang-Mills at large $N$ and large 't Hooft coupling.
We hope our techniques can be used and generalized to find more new sum rules from gravity.

We did not study the full current-current or stress tensor-stress tensor correlation functions, but only a single component of these multi-index objects.  Our fermionic Green's function turned out to be diagonal in spinor indices.  One would of course like to investigate the full current-current and stress tensor-stress tensor correlation functions, and also consider more generic cases where the fermionic Green's function is not diagonal.  Such cases would involve solving more general coupled systems of ordinary differential equations in gravity. Two distinct systematic approaches to such coupled systems 
were developed in \cite{Kaminski:2009dh,Buchel:2010gd}.

We leave the reader with the most interesting unanswered question of the paper: What is the physical meaning of the sum rule (\ref{genericsumrule}) in gravity?  Is there a dual gravitational interpretation of the sum rule itself?

\section*{Acknowledgments}
We thank Martin Ammon, Simon Caron-Huot, Sean Hartnoll, Carlos Hoyos, Patrick Kerner, Pavel Kovtun, Silviu Pufu, Jie Ren, and Antonello Scardicchio for discussion.
We thank the GGI for hospitality and the INFN for partial support during the completion of this work.
CH and DG were supported in part by the US NSF under Grants No.\
PHY-0844827 and PHY-0756966. CH thanks the Sloan Foundation for partial support. MK thanks the Erwin-Schr\"odinger-Institut (ESI) in Vienna for hospitality and partial support.
MK was supported in part by the DFG ({\it Deut\-sche For\-schungs\-ge\-mein\-schaft}).

\appendix

\section{The Holographic Conductivity Equation}
\label{app:conductivity}

In this appendix, we want to demonstrate how an equation of the form (\ref{eq:order2}) arises in computing a current-current correlation function from AdS/CFT.  We also review some simple cases where closed form expressions are available for these correlation functions.

Consider the action
\[
S =\int d^{d+1} x \sqrt{-g} \left[
 \frac{1}{2 \kappa^2} \left(R + \frac{d(d-1)}{L^2} \right) 
- \frac{1}{4 e^2} F_{ab}F^{ab}
- |D_a \Phi|^2- m^2 |\Phi|^2 \right]  \ ,
\]
where $D_a = \nabla_a - i A_a$.  
We look for a solution to the equations of motion 
with the ansatz $A = A_t(r) \, dt + A_x (t,r) \, dx$, $\Phi = \Phi(r)$ (with $\Phi$ a real function) and 
\begin{eqnarray}
\label{eq:metric}
ds^2 = - f(r) e^{-\chi(r)} dt^2 + \frac{r^2}{L^2} \, d \vec x^2  + \frac{dr^2}{f(r)} + 2 g_{tx}(t,r) \, dx \, dt \ .
\end{eqnarray}
We will treat the components $A_x$ and $g_{tx}$ as perturbations and only consider their linearized equations of motion.
As given in ref.\ \cite{Hartnoll:2008kx}, we find that the background satisfies the equations of motion
\begin{eqnarray}
\label{feq}
 \Phi'^2 + \frac{d-1}{2}\frac{f'}{\kappa^2 r f} + \frac{1}{2 e^2} \frac{e^\chi A_t'^2}{f}+ \frac{(d-1)(d-2)}{2\kappa^2 r^2}- \frac{d(d-1)}{2\kappa^2 L^2 f} +  && \nonumber \\
\qquad \qquad + \frac{e^\chi A_t^2 \Phi^2}{f^2}+ \frac{m^2 \Phi^2}{f} &=& 0 \ , \\
\label{chieq}
\frac{d-1}{2}  \frac{\chi'}{2 \kappa^2} + r \Phi'^2 + \frac{r e^\chi A_t^2 \Phi^2}{f^2} &=& 0 \ , \\
A_t'' + \left( \frac{d-1}{r} + \frac{\chi'}{2} \right) A_t' - \frac{ 2 e^2 A_t \Phi^2}{f} &=& 0 \ ,  \\
\Phi'' + \left( \frac{d-1}{r} +\frac{f'}{f} - \frac{\chi'}{2} \right) \Phi' + \left( \frac{ e^\chi A_t^2 }{f^2} - \frac{m^2}{f} \right) \Phi &=& 0 \ .
\end{eqnarray}
We impose the boundary conditions 
\begin{eqnarray}
f = \frac{r^2}{L^2} + \ldots \ , &&
e^{-\chi} f = \frac{r^2}{L^2}  + \ldots \ , \\
\Phi = O_\Delta r^{-\Delta} + \ldots \ ,  &&
A_t = \mu + \ldots \ ,
\end{eqnarray}
as $r \to \infty$ where $\Delta > (d-2)/2$ satisfies the unitarity bound.  We also want to assume the existence of a horizon at a radius $r=r_h$ where
$f(r_h) = 0$ and $A_t(r_h) = 0$ vanish linearly with $r$.  The other functions $\chi$ and $\Phi$ are finite at the horizon.  There is a relation between $A_t'(r_h) \equiv A_h$, $f'(r_h)\equiv f_h$, $\chi(r_h) \equiv \chi_h$ and $\Phi(r_h)\equiv \Phi_h$:
\begin{equation}
\frac{f_h}{r_h} = \frac{d}{L^2} - \frac{ \kappa^2}{d-1} \left( \frac{A_h^2 e^{\chi_h}}{e^2}
+ 2 m^2 \Phi_h^2 \right)  \ .
\end{equation}

Given such a black hole background, the linear fluctuations satisfy the differential equations
\begin{eqnarray}
g_{tx}'  - \frac{2}{r} g_{tx} + \frac{2 \kappa^2}{e^2} A_t' A_x &=& 0 \ , \\
A_x'' + \left( \frac{f'}{f} - \frac{\chi'}{2} + \frac{d-3}{r}\right) A_x' + \nonumber \\
\qquad \qquad + \left[ \left( \frac{\omega^2}{f^2}
- \frac{ 2 \kappa^2}{e^2} \frac{ A_t'^2}{ f} \right) e^{\chi} -
 \frac{2 e^2 \Phi^2}{f}
\right] A_x &=& 0 \ .
\label{Axeq}
\end{eqnarray}
This equation (\ref{Axeq}) is precisely of the general form (\ref{eq:order2}).

In the following subsections, we review the cases $d=2$, 3 and 4 when 
$\Phi = A_t = \chi=0$, and $f = r^2 ( 1 - (r_h/r)^d) / L^2$, and the equation 
(\ref{Axeq}) is exactly solvable.  One can locate the poles of the correlation function, deduce the large 
$\omega$ behavior, and verify the sum rules analytically.  
The sum rules for the corresponding correlation function $G_R^{xx}(\omega)$ are often rephrased in terms of the charge conductivity $\sigma(\omega) = G_R^{xx} (\omega) / i \omega$.

\subsection{Current-Current Correlators in Thermal Backgrounds}
\label{app:JJ}

We review some known and partially known results for current-current correlators calculated from dual AdS-Schwarzschild backgrounds.  
These are correlation functions in a thermal bath that preserves rotational symmetry.  The first part of our treatment follows \cite{Kovtun:2005ev}.  

Gauge invariance puts some constraints on the form of current-current correlation functions.  We expect that
\be
G^{\mu\nu}(\omega, \vec k) = P_{\mu\nu}^L\Pi^L(\omega, \vec k^2)+ P_{\mu\nu}^T \Pi^T(\omega, \vec k^2) \ ,
\ee
where
\be
P_{00}^T = 0 \ , \; \; \; P_{0i}^T = 0 \ , \; \; \; P_{ij}^T = \delta_{ij} - \frac{k_i k_j}{ \vec k^2} \ , 
\ee
\be
P_{\mu\nu}^L = P_{\mu\nu} - P_{\mu\nu}^T \ , \; \; \;
P_{\mu\nu} = \eta_{\mu\nu} - \frac{k_\mu k_\nu}{k^2} \ .
\ee
These correlation functions satisfy the gauge symmetry Ward identities because
 $k^\mu P_{\mu\nu}^T = 0 = k^\mu P_{\mu\nu}^L$.  

Our $d+1$ dimensional black brane metric is
\be
ds^2 = \frac{L^2}{z^2} \left( -f(z) dt^2 + d \vec x^2 + \frac{dz^2}{f(z)} \right) \ ,
\ee
where $f(z) = 1 - (z/z_h)^d$.   Consider the action for a Maxwell field in this background:
\be
S_{\rm EM} = - \frac{1}{4} \int d^{d+1}x \sqrt{-g} F_{ab} F^{ab} \ .
\ee 
We would like to consider fluctuations with a $e^{-i \omega t + i k x}$ dependence.   There are two types, longitudinal and transverse.  By a gauge choice, we set $A_z = 0$. The transverse fluctuations involve $A_a$ where $a \neq t$, $x$, or $z$, while the longitudinal fluctuations involve both $A_t$ and $A_x$.  
The transverse fluctuations satisfy the differential equation
\be
A_a'' + \left( \frac{f'}{f} + \frac{3-d}{z} \right) A_a' + \frac{\omega^2 - k^2 f}{f^2} A_a = 0 \ .
\label{Aaeq}
\ee
The longitudinal fluctuations satisfy the three coupled differential equations
\begin{eqnarray}
\omega A_t' + k f A_x' &=& 0 \ , \\
A_x'' + \left( \frac{f'}{f} + \frac{3-d}{z} \right) A_x' + \frac{ \omega^2}{f^2} A_x + \frac{ \omega k }{f^2} A_t &=& 0 \ , \\
A_t'' + \frac{3-d}{z} A_t' - \frac{ k^2}{f} A_t - \frac{k \omega}{f} A_x &=& 0 \ .
\end{eqnarray}
The third equation follows from the first two.  There is an equivalent ``gauge invariant'' formulation of this system of equations constructed from $E = k A_t + \omega A_x$:
\be
E'' + \left( \frac{\omega^2 f'}{f (\omega^2 - k^2 f) }+ \frac{3-d}{z}\right) E' + \frac{\omega^2 - k^2 f}{f^2} E = 0 \ .
\label{Eeq}
\ee

From solutions to the differential equations (\ref{Aaeq}) and (\ref{Eeq}), we can construct $\Pi^T$ and $\Pi^L$ respectively.  For a retarded Green's function, we choose ingoing boundary conditions at the event horizon: $E \sim A_a \sim (1-z)^{-i \omega z_h / d}$.  
At small $z$, $A_a$ and $E$ have the expansion
\be
a (1 + \ldots) + b z^{d-2} (1 + \ldots) \ ,
\ee
and the corresponding retarded Green's function, up to contact terms, should be $b / a$ for $d>2$.  For $d=2$, the expansion will be a little different, $b - a \ln z + \ldots$ (see for example 
\cite{Ren:2010ha}).

In the following subsections, we will work in units where $z_h = 1$.  Note that the temperature of the field theory is given by $T = d / 4 \pi z_h$.  To restore units, we shift $\omega \to \omega z_h = \omega d / 4 \pi T$.  

\subsection*{d=2}

In $d=2$ space-time dimensions, there are no transverse fluctuations, and we need solve only 
(\ref{Eeq}).  The solution with the correct ingoing boundary conditions at the event horizon can
be written in terms of hypergeometric functions \cite{Birmingham:2001pj}:
\begin{eqnarray}
E &=&  (1-z^2)^{-i \omega/2} 
\left[ (1-z^2) z^2 \left( \frac{\omega^2-k^2}{4} + i \omega - 1\right) \times
\right.   \\
&&
\times
{}_2 F_1 \left[ 2 - \frac{i (\omega+k)}{2}, 2 - \frac{i(\omega-k)}{2}, 2-i \omega; 1-z^2 \right] 
 \nonumber \\
&&
\left.
+ 
\left( \frac{2-i \omega}{2} z^2 - 1 \right) (i \omega-1)
{}_2 F_1\left[ 1 - \frac{i (\omega+k)}{2}, 1- \frac{i (\omega-k)}{2}, 1-i \omega; 1-z^2 \right]
\right] \ . \nonumber
\end{eqnarray}
(The other solution is the complex conjugate.)\footnote{%
 We would like to thank Jie Ren for discussions about this case.
}  
The near boundary expansion of this solution has the form $E = b - a \ln z$.  The corresponding Green's function is then
\be
\Pi^L(\omega, k) = 
-\frac{i \omega}{\omega^2 - k^2} - \frac{1}{2} \psi\left( \frac{i(k-\omega)}{2} \right) - \frac{1}{2} \psi\left(- \frac{i(k+\omega)}{2} \right) - \gamma+ \ln \Lambda  \ ,
\ee
where $\ln \Lambda$ is a renormalization dependent contact term and $\psi(x) = \Gamma'(x) / \Gamma(x)$. 
Note that $\Pi^L$ has poles in the lower half of the complex plane at positions $\omega = \pm k - 2 n i$ where $n = 0, 1, 2, \ldots$.  

The spectral function is 
\be
\rho(\omega, k ) = \mbox{Im} \, \Pi^L = \frac{\pi}{4} \left( 
\coth \left( \frac{ \pi}{2} (\omega+k) \right) + \coth \left( \frac{\pi}{2} (\omega-k) \right) \right) \ .
\ee
Since $\omega \rho(\omega, k)/ (\omega^2 - k^2) > 0$, it follows that $\omega \rho(\omega, k) P^L$ is a positive definite matrix, as it should be.  

There is a sum rule here of the form\footnote{%
 This integral is an example of Frullani's integral (which in turn is an application of Fubini's theorem).  Assuming that $df/dx = g(x)$, then
\be
 \int_0^\infty dx \, \frac{f(bx) - f(ax)}{x} = \int_a^b du \int_0^\infty dx \, g(ux) = \left( f(\infty) - f(0) \right) \ln (b/a) \ .
\ee
In our case, $f(x) \sim \coth(x) - 1/x$.
}
\be
\int_0^\infty \frac{d\omega}{\omega} \, \left( \rho(\omega/a,0) - \rho(\omega/b,0)  + \frac{b-a}{\omega}  \right) = \frac{\pi}{2} \ln \frac{b}{a} \ .
\label{frullani}
\ee
Thinking of $a$ and $b$ as shifting the temperature, we are regulating the sum rule by comparing the spectral density at two different temperatures and additionally subtracting off a divergent contribution at $\omega=0$.  Note that $\rho(\infty,0) = \pi/2$.
There are also a series of generalized sum rules of the form (\ref{gensumrule}):
\begin{eqnarray}
\frac{\psi^{(2)}(1)}{8} &=& \int \frac{ d\omega}{\pi \omega^3} 
\left( \rho(\omega,0) - \frac{1}{\omega} - \frac{\pi^2 \omega}{12} \right) \ , \nonumber \\
-\frac{\psi^{(4)}(1)}{384} &=& \int \frac{ d\omega}{\pi \omega^5} 
\left( \rho(\omega,0) - \frac{1}{\omega} - \frac{\pi^2 \omega}{12} + \frac{\pi^4 \omega^3}{720} \right) \ , \nonumber \\
\frac{\psi^{(6)}(1)}{46{,}080} &=& \int \frac{ d\omega}{\pi \omega^7} 
\left( \rho(\omega,0) - \frac{1}{\omega} - \frac{\pi^2 \omega}{12} + \frac{\pi^4 \omega^3}{720} 
- \frac{\pi^6 \omega^5}{30{,}240} \right) \ , \nonumber \\
&\vdots& \nonumber \\
\frac{\psi^{(2n)}(1)}{(4n)!!} (-1)^{n+1} &=& \int \frac{d\omega}{\pi \omega^{2n+1}}
\left(
\rho(\omega,0) - \frac{1}{\omega} - \frac{1}{\omega} \sum_{k=1}^n \frac{
(\pi \omega)^{2k} 
B_{2k}
}{(2k)!} \right) \ .
\label{d2gensum}
\end{eqnarray}
The subtractions here regulate the IR divergence at $\omega=0$.

\subsection*{d=3}

For $d>2$, an exact solution is available for (\ref{Aaeq}) and (\ref{Eeq}) only in the case $k=0$.  Note that when $k=0$, the two equations are identical and $\Pi^L(\omega,0) = \Pi^T(\omega, 0)$.  

For $d=3$, the authors of \cite{Herzog:2007ij} found the solution
\be
A_a(z) = E(z) = \exp \left( i \omega \int_0^z \frac{dy}{f(y)}\right) \ .
\ee
The corresponding Green's function is
\be
\Pi(\omega, 0) = i \rho = i \omega \ .
\ee
Note that the Green's function in this case has no poles.
The spectral function satisfies the trivial sum rule
\be
\int_{-\infty}^\infty \left( \frac{\rho}{\omega} - 1 \right) d \omega = 0 \ .
\ee

\subsection*{d=4}
\label{app:d4}

For $d=4$, the authors of \cite{Myers:2007we} found the solution 
\begin{eqnarray}
E &=& \left( 1 - \frac{1}{z^2} \right)^{-i \omega/4} \left( 1 + \frac{1}{z^2} \right)^{-\omega/4}
\times \\
&&
\qquad \qquad \qquad
{}_2 F_1 \left[
1 - \frac{1+i}{4}\omega, - \frac{1+i}{4} \omega, 1 - \frac{i\omega}{2}, \frac{1}{2} \left(1-\frac{1}{z^2} \right)
\right] \ . \nonumber
\end{eqnarray}
(The other solution is again the complex conjugate.)
The Green's function is
\be
\Pi(\omega,0) = - i \omega  -\frac{\omega^2}{2} \left[  2 \gamma + \ln \Lambda' 
+ \psi \left( - \frac{1+i}{4} \omega \right) + \psi \left( \frac{1-i}{4} \omega \right) \right] \ .
\ee
This Green's function has poles in the lower half of the complex plane at positions
$\omega = 2n (\pm 1 -i)$ where $n=0,1,2,\ldots$. 
For comparison with a result in the body of the paper, we give the large $\omega$ expansion of 
$\Pi(\omega,0)$:
\be
\Pi(\omega,0 ) = 
\omega^2 \ln (i \Lambda / \omega)
+ \frac{8}{15 \omega^2} 
+ O(\omega^{-4}) \ .
\label{largeomegad4JxJx}
\ee
The spectral density is
\begin{eqnarray}
\rho 
&=&
\frac{\pi \omega^2 \sinh (\pi \omega/2)}{2 (\cosh (\pi \omega/2) - \cos(\pi \omega/2) )} \ .
\end{eqnarray}
An indefinite integral is known
\begin{eqnarray}
\int \frac{\rho}{\omega} d \omega &=&
\frac{\pi \omega^2}{4} + \frac{\omega}{2} (1-i) \ln \left( 1- e^{- \pi \omega(1+i)/2} \right)
+\frac{\omega}{2}(1+i) \ln \left( 1 - e^{-\pi \omega (1-i)/2} \right) \nonumber \\
&&
+ \frac{i}{\pi} \mbox{Li}_2 \left( e^{-\pi \omega(1+i)/2} \right)
- \frac{i}{\pi} \mbox{Li}_2 \left( e^{-\pi \omega (1-i)/2} \right) \ .
\end{eqnarray}
From this integral, it is straightforward to verify the sum rule
\be
\int_0^\infty \frac{1}{\omega} \left( \rho - \frac{\pi \omega^2}{2} \right) d \omega = 0 \ .
\ee
This integral was studied numerically by \cite{Baier:2009zy}.

Like in the $d=2$ case, there are some more interesting sum rules of the form (\ref{gensumrule}) that one can write down:
we have
\begin{eqnarray}
-\frac{\zeta(5)}{2^6} &=& \int \frac{d \omega}{\pi \omega^7} \left(
\rho - \omega - \frac{\pi^2 \omega^3}{24} - \frac{\pi^4 \omega^5}{2880} \right) \ , \nonumber \\
\frac{\zeta(9)}{2^{12}} &=& 
\int \frac{d \omega}{\pi \omega^{11}} \left(
\rho - \omega - \frac{\pi^2 \omega^3}{24} - \frac{\pi^4 \omega^5}{2880} + \frac{\pi^6 \omega^7}{241{,}920}
+ \frac{\pi^8 \omega^9}{19{,}353{,}600}
\right) \ \ , \nonumber \\
&\vdots& \nonumber \\
\frac{\zeta(4n+1)}{2^{6n}} (-1)^n &=&
\int \frac{d \omega}{\pi \omega^{4n+3}}
\Bigl( \rho - \omega + \omega \sum_{k=1}^n (-1)^k  
\left( \frac{ \pi^2 \omega^{2}}{2} \right)^{2k} \times
\nonumber \\
&&
\hskip1in \left( \frac{ 2 B_{4k-2}}{\pi^2 \omega^2 (4k-2)!  }
- \frac{B_{4k}}{(4k)! } \right) \Bigr) \ .
\label{d4gensum}
\end{eqnarray}
The subtractions regulate the IR divergence at $\omega=0$.

\section{The Holographic $T^{xy}$ Two Point Function}
\label{app:TT}

In this appendix, we review how an equation of the form (\ref{eq:order2}) arises in computing a stress-tensor two-point function from AdS/CFT.
Consider the action
\be
S = \frac{1}{2 \kappa^2} \int d^{d+1} x \sqrt{-g} \left( R + \frac{d(d-1)}{L^2} \right) 
+ \frac{1}{\kappa^2} \int d^d x \sqrt{-g_{\rm bry}} \left( K - \frac{d-1}{L} \right) \ ,
\ee
where $K = \nabla^\mu n_\mu$ (sum on only the gauge theory indices) is the trace of the extrinsic curvature and $n_\mu$ is a unit vector normal to the boundary and pointing toward larger $r$.  We take the ansatz for the metric
\be
ds^2 = - f(r) e^{-\chi(r)} dt^2 + \frac{r^2}{L^2} d \vec x^2 + \frac{dr^2}{f(r)} + 2 \frac{r^2}{L^2} \phi(r,t,z) \, dx \, dy \ .
\ee
The tensor $g_{\rm bry}$ is the induced metric on a constant $r$ slice.  

We define two quantities, the bulk Lagrangian and the zeroth order Einstein tensor:
\be
{\mathcal L} \equiv \frac{\sqrt{-g}}{2 \kappa^2} \left( R + \frac{d(d-1)}{L^2} \right) \ , \; \; \;
{G^a}_b \equiv \left. \frac{\delta S}{\delta {g_a}^b} \right|_{\phi = 0} \ .
\ee
The first observation is that
\begin{eqnarray}
\label{bigsubtract}
\mathcal L - (2-\phi^2) {G^x}_x  &=& 
-\frac{1}{4 \kappa^2} \sum_{a = t,r,z} \left[ \sqrt{-g} (\partial_a \phi)( \partial^a \phi) 
- 4 \partial_a (  \sqrt{-g}  \, \phi \partial^a \phi) \right] 
\nonumber \\
&&
- \frac{1}{\kappa^2} \partial_r \left( \frac{\sqrt{-g} f}{r}
\right)
+ O(\phi^3) \ .
\end{eqnarray}
Note that this equality is only valid to order $\phi^3$.
From this observation, because of the $(\partial \phi)^2$ term in the Lagrangian, we conclude that (\ref{eq:order2}) does indeed govern the stress-tensor two-point function.
Also, we find the onshell action reduces to 
\begin{eqnarray}
S_{\rm os} &=& \frac{1}{\kappa^2} \int d^dx  \sqrt{-g_{\rm bry}} \left( K - \frac{d-1}{L} -
n^r \left( \frac{1}{r} 
- \frac{3}{4} \phi \partial_r \phi \right) \right) \ .
\label{Sos}
\end{eqnarray}
With the addition of matter fields to the action, one should include the bulk stress tensor in making the subtraction (\ref{bigsubtract}).  However, provided these matter fields have high enough scaling dimension, they will not affect (\ref{Sos}) in the large $r$ limit.

Consider the simple case of a constant metric perturbation where $\phi = \phi_0$ is a constant. 
We will assume that for large $r$, $\chi \sim r^{-2 \Delta}$ where $\Delta > d/2$ and that
$f \sim (r^2/L^2) (1 - 2 \kappa^2 L^{d+1} p / r^d + \ldots)$.  In this case, the on-shell action 
evaluates to
\be
S_{\rm os} = \int d^d x \, p \left( 1 - \frac{1}{2} \phi_0^2 \right) \ .
\ee
We conclude two things.  The constant $p$ can be interpreted as the pressure and 
$G_R^{xy,xy}(0,0) = -p$.  

To see what happens when $\phi$ is space-time dependent, let's consider 
the specific cases $d=3$ and $d=4$.  Again, this analysis should generalize  
to the case where there are additional matter fields in the action, provided the matter fields have conformal scaling dimension $\Delta > d/2$.

\subsection*{d=4}

Near the boundary, $\phi$ has the expansion
\be
\phi = \phi_0 + \frac{ \phi_0 L^4 (\omega^2 - k^2)}{4 r^2} + \frac{\phi_1}{r^4}
+ \frac{\phi_0 L^8 ( \omega^2 - k^2)^2}{16 r^4} \ln r + \ldots 
\ee
This expansion yields the on-shell action
\begin{eqnarray}
S_{\rm os} &=& \int d^4 x \left[ \frac{\omega^2 - k^2}{8 L \kappa^2} \phi_0^2 \, r^2 
+ p \left( 1 - \frac{1}{2} \phi_0^2 \right) \right.
\\
&&
\left. +
\frac{\phi_0 \phi_1}{L^5 \kappa^2} 
+ \frac{L^3}{64 \kappa^2} (k^2 - \omega^2)^2 ( 1 + 4 \ln r)\phi_0^2  + O(r^{-1})
\right]
\ . 
\nonumber
\end{eqnarray}

To regulate the $\ln(r)$ and $r^2$ divergences, we can add boundary counter-terms
\be
S_{\rm ct} = \frac{L}{ 4 \kappa^2} \int d^4x \sqrt{-g_{\rm bry}} \, 
\left(
 {\mathcal R} - 
 \frac{ L^2 \ln \Lambda'}{2} \, {\mathcal R}^{\mu\nu} {\mathcal R}_{\mu\nu}
 \right) \ ,
\ee
where ${\mathcal R}_{\mu\nu}$ and ${\mathcal R}$ are the boundary Ricci tensor and Ricci scalar respectively.  At leading order in $\phi$,
\begin{eqnarray}
{\mathcal R} &=& \frac{1}{2} \left( \frac{ L^2 k^2}{r^2} - \frac{ \omega^2 e^{\chi}}{f} \right) \phi^2 + O(\phi^3) \ , \\
{\mathcal R}^{\mu\nu} {\mathcal R}_{\mu\nu}
&=& \frac{1}{2} \left( \frac{L^2 k^2}{r^2} - \frac{\omega^2 e^{\chi}}{f} \right)^2 \phi^2 + O(\phi^3) \ .
\end{eqnarray}
The end result is that
\be
S_{\rm os} + S_{\rm ct} = \int d^4x \left[ p \left(1 - \frac{1}{2} \phi_0^2 \right) + \frac{\phi_0 \phi_1}{L^5 \kappa^2} + \frac{ L^3 \phi_0^2}{16 \kappa^2} (k^2 - \omega^2)^2 \ln \Lambda
\right] \ .
\ee

\subsection*{d=3}

Near the boundary, $\phi$ has the expansion 
\be
\phi = \phi_0 + \frac{L^4 \omega^2}{2 r^2} \phi_0 + \frac{\phi_1}{r^3}
+\ldots
\ee
This expansion yields the on-shell action
\be
S_{\rm os} = \int d^3 x \left[ \frac{\omega^2 }{4 \kappa^2} \phi_0^2 \, r 
+ p \left( 1 - \frac{1}{2} \phi_0^2 \right) 
+ \frac{3}{4 L^4 \kappa^2} \phi_0 \phi_1
\right]
\ . 
\ee

To regulate the $r$ divergence, we can add a boundary counter-term
\be
S_{\rm ct} = \frac{L}{ 2 \kappa^2} \int d^4x \sqrt{-g_{\rm bry}} \, 
 {\mathcal R} \ ,
\ee
where ${\mathcal R}$ is the boundary Ricci scalar.  At leading order in $\phi$,
\be
{\mathcal R} = -\frac{1}{2}
\frac{e^{\chi} \omega^2}{f}
 \phi^2 + \ldots
\ee
The end result is that
\be
S_{\rm os} + S_{\rm ct} = \int d^3x \left[ p \left(1 - \frac{1}{2} \phi_0^2 \right) + \frac{3\phi_0 \phi_1}{4 L^4 \kappa^2} \right] \ .
\ee

\section{Contraction Maps}
\label{app:contract}

\subsection{Bosons}
\label{app:contractbosons}

We prove
\be
\mathcal{I}(s) = \int_{r_H}^r \mathcal{E}(s(r')) dr' \ 
\ee
is a contraction mapping.
First, we restrict the set of $s$ that ${\mathcal I}$ acts on to a domain $D = \{ s : || s||_\infty < K \}$
for some constant $K$ where $||s||_\infty \equiv \sup_{r\in [r_H, \infty[} | e^{-2 i \omega u} s(r) |$
We will show first that ${\mathcal I}(s) \in D$.  Then, 
we will show that for sufficiently large $\omega$, $\mathcal{I}$ is a
contraction mapping with contraction factor $O(|\omega|^{-1})$.  More specifically, we show that
\be \label{eq:contractionIneq}
\left | \mathcal{I}(s)-\mathcal{I}(s_0)\right | \leq \frac{c}{|\omega|} || s - s_0 ||_\infty \qquad \forall \, s,\, s_0 \,\in D\, ,
\ee 

Note that
\begin{eqnarray}
\label{Isineq}
| {\mathcal I}(s) | & \leq &
|| s  ||_\infty^2 \int_{r_h}^\infty  dr' \frac{|e^{4 i \omega u} P {\mathcal D}(P)|}{|PQ'-P'Q|} \\
&&
+ || s ||_\infty \int_{r_h}^\infty \frac{|e^{2 i \omega u} P {\mathcal D}(Q)| + 
| e^{2 i \omega u} Q {\mathcal D}(P)|}{|PQ'-P'Q|} 
+ \int_{r_h}^\infty \frac{ | Q {\mathcal D}(Q) | }{|PQ'-P'Q|} \nonumber \ .
\end{eqnarray}
We will show below that each of these four integrals scales as $1/|\omega|$ in the large $\omega$
limit, provided $\mbox{Im} \, \omega \geq 0$:
\begin{eqnarray*}
\int_{r_h}^\infty  dr' \frac{|e^{4 i \omega u} P {\mathcal D}(P)|}{|PQ'-P'Q|}  \le
 \frac{C_{PP}}{|\omega|} &,&
 \int_{r_h}^\infty  dr' \frac{|e^{2 i \omega u} P {\mathcal D}(Q)|}{|PQ'-P'Q|}  \le
 \frac{C_{PQ}}{|\omega|} ,\\
\int_{r_h}^\infty  dr' \frac{|e^{2 i \omega u} Q {\mathcal D}(P)|}{|PQ'-P'Q|}  \le 
 \frac{C_{QP}}{|\omega|} &,&
\int_{r_h}^\infty  dr' \frac{| Q {\mathcal D}(Q)|}{|PQ'-P'Q|}  \le
 \frac{C_{QQ}}{|\omega|} \ .
\end{eqnarray*}
The inequality (\ref{Isineq}) reduces to
\be
| {\mathcal I}(s) | \le \frac{1}{|\omega|} \left( K^2 C_{PP} + K (C_{PQ} + C_{QP}) + C_{QQ} \right) \ .
\ee
Thus if $K$ is of order one compared to $\omega$, then ${\mathcal I}(s) \in D$.

Next we evaluate the l.h.s. of equation \eqref{eq:contractionIneq}
\begin{eqnarray}
\left | \mathcal{I}(s)-\mathcal{I}(s_0)\right |  &\leq& 
|| s - s_0 ||_\infty  \left( || s + s_0 ||_\infty 
\int_{r_h}^\infty  dr' \left| \frac{ e^{4i \omega u}  P \mathcal{D}(P)}{PQ'-P'Q} \right|
+ \right. \\
&& \hspace{25mm} \left. +
 \int_{r_h}^\infty  dr' \frac{|e^{2 i \omega u} (P {\mathcal D}(Q) + Q {\mathcal D}(P))|}{|PQ'-P'Q|} 
\right)  \nonumber \\
&\le & \frac{\left( 2 K C_{PP} + C_{PQ} + C_{QP} \right) }{|\omega|}   || s - s_0 ||_\infty \ .
\end{eqnarray}
is a contraction mapping with $ c = 2 K C_{PP} + C_{PQ} + C_{QP}$.

\subsection*{A Lemma}

To demonstrate that
\be
\int_{r_H}^{\infty} \left|  \frac{e^{4i \omega u} P \mathcal{D}(P)}{PQ'-P'Q} \right| dr = 
\int_0^{\infty} \left|  \frac{e^{4i \omega u} P \mathcal{D}(P)}{PQ'-P'Q} \right| F du \leq \frac{C_{PP} }{|\omega|} \ ,
\ee 
we first perform a change of variables so that the integral is over $u$.
Then we split the integral. 
The Hankel functions $H_{(n+1)/2}^{(i)} (x)$ have the property that
$|H_{(n+1)/2}^{(i)}(x)|=O(x^{-1/2})$ as $x \rightarrow \infty$,
$|H_{(n+1)/2}^{(i)} (x)|=O(x^{-(n+1)/2})$ as $x \rightarrow 0$.
The quantity $y(u)$
has the property that
$|y(u)|=O(u^{-1})$ as $u \rightarrow \infty$. 
We assume that $y(u)=O(u^{2\Delta-1})$ as $u \rightarrow 0$ for some $\Delta$.
We split the integral into three parts according to whether
$u$ is greater or smaller than some fixed value $u_0$, and according to whether
$|\omega| u$ (the absolute value of  the 
argument of the Hankel function) is greater or smaller than one.
So the three regions are $(u_0,\infty)$, $(|\omega|^{-1},u_0)$, and
$(0,|\omega|^{-1})$.

On $(u_0,\infty)$ we get a bound of $\int_{u_0}^{\infty} O((|\omega| u)^{-1}) O(u^{-1}) du = O(|\omega|^{-1})$.
On $(|\omega|^{-1},u_0)$ we get a bound of
$\int_{|\omega|^{-1}}^{u_0} O((|\omega| u)^{-1}) O(u^{2\Delta-1}) du$,
which is $O(|\omega|^{-1})$ if $2\Delta > 1$.
On $(0,|\omega|^{-1})$ we get a bound of $\int_0^{|\omega|^{-1}} O((|\omega| u)^{-n-1}) O(u^{2\Delta-1}) du$, which is $O(|\omega|^{-2\Delta})$ if $2\Delta > n+1$.
So, for $n \ge 0$, $2\Delta > n+1$, the integral converges and is
$O(|\omega|^{-1})$ for large $\omega$.

For the current-current correlator, we have
$n=d-3$, so we have convergence if $2\Delta > d-2$.  For the stress tensor
correlator, $n=d-1$, so we have convergence if $2\Delta > d$.

The result for $C_{PQ}$, $C_{QP}$, and $C_{QQ}$ and 
the other three integrals follow analogously.
So, assuming $\Delta$ is sufficiently large, $s \mapsto \int \mathcal{E}(s)$
is a contraction mapping with contraction factor $O(|\omega|^{-1})$.
We can use this mapping to find an asymptotic expansion in powers of
$|\omega|^{-1}$ for the solution to (\ref{eq:s}).

\subsection{Fermions}
\label{app:contractfermions}

We show that ${\mathcal I}(s)$ \eqref{eq:I} is a contraction-like mapping.
(We assume $\mbox{Im} \, \omega \geq 0$.) 
\begin{eqnarray}
\psi^{(2)}&= & \mathcal{I}(\mathcal{I}(\mathcal{I}(0))) - \mathcal{I}(\mathcal{I}(0))\, , \\
 && \vdots \nonumber \\
\psi^{(j)}&= & \II^{\circ (j+1)}(0) - \II^{\circ j}(0) \label{eq:fermionCorrections}
\end{eqnarray}
where $\II^{\circ j}(0) = \underbrace{\II(\II(...\II}_j(0)))$.
Equivalently, $\II^{\circ j}(0) = \psi^{(0)}+\psi^{(1)}+...+\psi^{(j-1)}$.
These contributions are the corrections, which
extend the solution $\psi^{(0)}$ to nonzero momentum $k$ and introduce a non-vanishing
scalar $\Phi$. If $\sum_{i=0}^{\infty} {\psi}^{(i)}$ converges,
then it converges to a solution of (\ref{eq:fermion}).

We would like to show that $\psi^{(j)}$ decreases exponentially with $j$,
which would imply that the sum converges.
First, we observe from the definitions of $\II$ and $\JJ$ that
$\II(\psi_1)-\II(\psi_2)=\JJ(\psi_1)-\JJ(\psi_2) = \JJ(\psi_1 - \psi_2)$.
We apply this identity to the right hand side of (\ref{eq:fermionCorrections}) to obtain
\begin{equation}
\psi^{(j)} = \JJ(\II^{\circ j}(0) - \II^{\circ (j-1)}(0)) = \JJ(\psi^{(j-1)}).
\end{equation}
It turns out that $\JJ(\psi)$ is not necessarily smaller than $\psi$.
However, we can still show that
$\JJ(\JJ(\psi))$ is smaller than $\psi$ by a factor of $O(\omega^{-\min(1,\Delta)})$,
where $\Delta$ is the scaling dimension of $\Phi$.  Given our assumption $\Delta > 1$,
we have that $\min(1, \Delta) = 1$.  
We write
\begin{eqnarray}
 \mathcal{J}(\mathcal{J}(\psi)) &= &
 -\int\limits_u^\infty du'
 \int\limits_{u}^{u'} du''
 \left [ 
 e^{-i\gamma^{\ur} \gamma^{\ut} (\omega (u''-u)+v(u'')-v(u))} \gamma^{\ur} 
 \left .\left(\sqrt{g^{xx}}k \gamma^{\ux} +i \Phi \right) \sqrt{-g_{tt}} \right |_{u''}
 \right ] \nonumber \\
& & \times \left [ 
 e^{-i\gamma^{\ur} \gamma^{\ut} (\omega (u'-u'')+v(u')-v(u''))} \gamma^{\ur} 
 \left. \left(\sqrt{g^{xx}}k \gamma^{\ux} +i \Phi \right)\sqrt{-g_{tt}}\right |_{u'}
\right ] \psi  \label{eq:jj} \\
&=&  -\int\limits_u^\infty du' \mathcal{J}(u,u')\, \psi \, , \nonumber \\ \label{eq:J}
\end{eqnarray}
where we have changed the order of integration and defined a new function
$\mathcal{J}(u,u')$.

\subsection*{Another Lemma}
We now examine the properties of $\mathcal{J}(u,u')$ defined in equation \eqref{eq:J}
in order to prove the fact that the corrections converge, i.e. that 
\be
\|{\psi}^{(i)}\| \le C |\omega|^{-1} \| \psi^{(i-2)}\|
\ee
is satisfied for some $C$, where we define
$\|\psi\| = \sup |e^{i (-\omega u + v)} \psi |$. 

Consider 
\begin{eqnarray}
\mathcal{J}(u,u') 
&= &
e^{i \gamma^{\ur} \gamma^{\ut} (\omega (u+u') + (v(u)+v(u'))}
\left.
	\left[ \left(
		\sqrt{g^{xx}}k \gamma^{\ux} +i
		\Phi
	\right) \sqrt{-g_{tt}} \right]
\right |_{u'} \nonumber \\
& & \cdot \int\limits_{u}^{u'} du'' 
e^{-2 i\gamma^{\ur} \gamma^{\ut} (\omega u''+ v(u''))} 
\left.
	\left[ \left(
		-\sqrt{g^{xx}}k \gamma^{\ux} +i
		\Phi
	\right) \sqrt{-g_{tt}} \right]
\right |_{u''} \label{eq:jint}
\end{eqnarray}
Once again we have a rapidly oscillating integral.
We can bound the integral using integration by parts.
Let $f$ be the integrand in \eqref{eq:jint}.
\begin{eqnarray}
\left| \int_u^{u'} e^{i \omega u''} f(u'') du'' \right| & = &
\frac{1}{|\omega|} \left| -i e^{i \omega u''} f(u'')|_u^{u'} + i \int_u^{u''}
e^{i \omega u''} f'(u'') du'' \right| \nonumber \\ \label{eq:fbound}
& \le & |\omega|^{-1} \left( 2 \sup |f(u'')| + \int_0^{\infty} |f'(u'')| du'' \right) \nonumber \\
& = & O(|\omega|^{-1}).
\end{eqnarray}
since $f$ and $f'$ decrease exponentially as $u$ becomes large.
The quantity on the first line of \eqref{eq:jint} decreases exponentially as
$u'$ becomes large.
Then
\begin{eqnarray}
\int_u^{\infty} J(u,u') \psi(u') du' & = & \int_u^{\infty} O(|\omega|^{-1} e^{-u'}) \psi(u')  du' \nonumber \\
& = & O(|\omega|^{-1}) \|\psi\|.
\end{eqnarray}
We have thus shown that $\sum\limits_j^\infty \psi^{(j)}$ converges to a
solution of the equation of motion \eqref{eq:fermion}.

\section{On the Absence of Branch Cuts}
\label{app:branchCutsGeneral}
In field theory, retarded Green's functions typically have branch cuts in addition to poles in the lower half of the complex frequency plane.  Because of our reliance on classical gravity and backgrounds with non-extremal black hole horizons, the expectation is that the field theories are in a large $N$ strong coupling 
limit and at nonzero temperature.
In this case, the common lore is that the Green's functions will have only poles \cite{Hartnoll:2005ju,Kovtun:2003vj}.  
Indeed, the exact results (at $T \neq 0$) we find in Appendix \ref{app:JJ} lack branch cuts.\footnote{%
  In regulating the large frequency behavior of these exact Green's functions, we often add logarithmic terms.  However, we are free to choose the branch cut of the regulator 
to lie in the lower half plane.
}  

On the gravity side, $G_R(\omega,k)$ is free of branch cuts in the upper half of the 
complex frequency plane essentially because the differential equation (\ref{eq:order2}) is holomorphic in $\omega$.  
In the next section, we argue that given an assumption about the singular points of the differential equation (\ref{eq:order2}), the singularities of $G_R(\omega,k)$, possibly away from a set of discrete points on the negative imaginary axis, are entirely determined by the quasinormal mode solutions we studied above.  
The assumption is the  requirement that  $r=r_h$ and $r=\infty$ be regular singular points of (\ref{eq:order2}), and that there be no other ``nearby'' regular singular points.  More specifically, we want the Frobenius power series solutions at $r=r_h$ and $r=\infty$ to have an overlapping region of validity along the real line $r_h  < r < \infty$.\footnote{%
 This assumption can probably be weakened and the following argument made to work provided the ``nearby'' singular points are regular and their indicial exponents independent of $\omega$.
 }
 
We repeat our analysis for fermionic Green's functions in Appendix \ref{app:branchCutsFermions}. In Appendix \ref{app:extension}, we outline a method that will allow for more general types of singularities in the differential equation.  This second method does not rule out branch cuts in the lower half plane, and may be useful for studying gravity systems away from the large $N$ and strong coupling limit.

\subsection{Bosons}
\label{app:branchCuts}

Given our assumption, there exists a Frobenius series solution to (\ref{eq:order2}) at 
$r=\infty$ which is holomorphic in $\omega$ on the interval $r_h < r < \infty$.  
The series solution at $r=\infty$ will generically take the form:
\be
A(r) = a  r^{\Delta - n-1} \sum_{j=0}^\infty a_j r^{-j} + b  r^{-\Delta} \sum_{j=0}^\infty b_j r^{-j} 
\label{bryseries}
\ee
where $-\Delta$ and $\Delta-n-1$ are called indicial exponents and satisfy the quadratic relation
$\Delta (\Delta - n - 1) = m^2 L^2$.  (We are allowing $Y$ to include a mass term.)  
The coefficients 
$a$ and $b$ are arbitrary and independent of $r$, and we choose $a_0 = b_0 = 1$.  
The fact that (\ref{bryseries}) is holomorphic in $\omega$ is perhaps obvious:
The indicial exponents of the series do not depend on $\omega$, and therefore $a_j$ and $b_j$ will be polynomials in $\omega$.   
Note that if $2 \Delta - n-1$ is an integer, the expansion may be modified to include logarithms.
But the logarithms are functions of $r$, not $\omega$.  

The existence of a holomorphic series solution at $r=r_h$ is less obvious as the indicial exponents depend on $\omega$.  We have:
\be
A_\pm(r) = (r-r_h)^{\pm i \omega / F_h} \sum_{j=0}^\infty c_{\pm j} (r-r_h)^j \ .
\label{horseries}
\ee
We assume $c_{\pm 0} = 1$.
The coefficients of the ingoing Frobenius series $A_-(r)$ 
will involve poles at regular intervals $2 i \omega / F_h = \mathfrak{n}$ where $\mathfrak{n}$ is a positive integer.
The poles come from the fact that for these values of $\omega$, the two horizon power series  $A_\pm(r)$ solutions overlap and $A_-$ should be modified to include $\ln (r-r_h)$ dependence. 
There is a similar problem with $\omega=0$.  Nonetheless, we can conclude that $A_-(r)$ is holomorphic away from these particular values of $\omega$ and that moreover $A_-(r)$ is holomorphic in the upper half of the complex frequency plane.

To determine $a$ and $b$ in (\ref{bryseries}), and hence the Green's function, we can match the ingoing Frobenius power series at the horizon (\ref{horseries}) (and its derivative) to the one at the boundary (\ref{bryseries}) at some intermediate point $r_h < r< \infty$.  The matching involves solving a linear system of two equations and two unknowns, $a$ and $b$.  Because there must exist a globally well defined solution, the determinant of the system cannot vanish, and the resulting $a$ and $b$ must be holomorphic in $\omega$, away from the special values mentioned in the previous paragraph.\footnote{%
The determinant is the Wronskian for the two power series solutions in (\ref{bryseries}) for which $W_f = (2 \Delta - n-1) r^{-n}/F(r)$.  
Superficially then we appear to have missed $2 \Delta = n+1$.  In this case, an extra logarithm in one of the series expansion guarantees that $W_f$ does not vanish.
}

The Green's function can be computed from the near boundary limit of $A'/A$ where we know $a$ and $b$ are holomorphic in the upper half plane and at worst meromorphic in the lower half plane.  
Hence, any branch cut or singular behavior of the Green's function 
in the upper half plane 
must come from zeroes of $a$, which we studied and largely ruled out, except at $\omega =0$, in the previous subsections.

If we relax any of the three initial assumptions of large $N$, strong coupling, or $T\neq 0$, the argument above generally ceases to work.
So, for example there is a possible source of confusion regarding the $T=0$ limit where we know that the Green's functions do often have branch cuts.  From the gravity side, the regular singular point at $r=r_h$ can become an irregular singular point at $r=0$ in the limit $T\to 0$.  The asymptotic series expansions around irregular singular points typically have a zero radius of convergence, and the matching argument above fails. 
Similarly, $1/N$ corrections and finite coupling are both known individually to introduce branch cuts \cite{Hartnoll:2005ju,Kovtun:2003vj}
in the lower half of the frequency plane.
The technique outlined in  Appendix \ref{app:extension} may be useful for these cases although it does not work for $T=0$.

\subsection{Fermions}
\label{app:branchCutsFermions}

Similar to the discussion of branch cuts in the bosonic case of Appendix \ref{app:branchCuts},
we argue that there are no branch cuts in our  fermionic Green's functions away from $T=0$. 
Instead of working with a second order differential equation, we find it simpler to work directly with the first order system \eqref{Diraceqsimp}.

Comparing with the bosonic case of Section \ref{app:branchCuts}, 
we make the same assumptions about the regular singular points of the differential equation, and we attempt to match a power series solution at the horizon to a power series solution at the boundary.  
The fermion 
equation of motion \eqref{Diraceqsimp} may be written
\be
\label{eq:schematicFermion}
\psi'_\alpha = \mathfrak{T} \psi_\alpha \, 
\ee
where $\mathfrak{T}$ is a $2\times 2$ matrix encoding the coefficients from \eqref{Diraceqsimp}.
In computing these power series, diagonalizing the matrix $\mathfrak{T}$
at the regular singular points may introduce square root branch cuts.  In fact, the boundary and horizon power series expansions are free of such cuts.  For the massless fermion, the boundary is actually not singular, and no diagonalization is required.
Near the horizon, we know
\be
\mathfrak{T}\sim -i \frac{\omega}{(r-r_h)F_h}
\left(
\begin{array}{c c}
0 & 1\\
1 & 0
\end{array}
\right)\, .
\ee
The eigenvalues and eigenvectors of  
$ \lim_{r\to r_h} (r-r_h)\mathfrak{T}$ have no square root branch cuts.  The Frobenius power series for the infalling solution will have at most logarithms and only for the same special values of $\omega$ along the negative frequency axis that we found in the bosonic case.
Thus the horizon and boundary expansion of $\psi$ are holomorphic in $\omega$ everywhere except for possibly a discrete set of $\omega$ along the negative frequency axis.  We can conclude the only singularities in the fermionic Green's function (away from these discrete values) come from quasinormal modes.
It should be possible to relax the assumptions about the regular singular points of \eqref{Diraceqsimp}
along the lines of Appendix \ref{app:extension}.  

\subsection{An Extended Class of Models }
\label{app:extension}

We outline a refined technique for demonstrating the absence of branch cuts in the upper half of the complex plane; we consider
only $\mathrm{Im}\, \omega\ge 0$ and $\omega \neq 0$.
In proving the absence of branch cuts before, recall that we restricted our differential equations to have a very limited class of singular behavior in Section \ref{app:branchCuts} and \ref{app:branchCutsFermions}.  
We believe this restriction is consistent with field theories at $T\neq 0$, large $N$ and strong 
coupling. 
The technique here is valid for an extended class of models where the Green's functions come from solving differential equations with worse types of singularities. 
Such singularities may occur in moving away from the large $N$ and strong coupling
limit in field theory.  However, this particular technique will fail at $T=0$.

We make the following assumptions:
\begin{itemize}
\item $\frac{1}{F}$ and $Y$ are integrable on the interval $(r_H,\infty)$, and $F$ is positive.
\item Near the horizon, $F \sim r-r_H$ and the integral of $Y$ converges.
\item Near the boundary, $F \sim r^2$ and $Y = O(r^{-2}).$
\end{itemize}

Let $s = \frac{-iFr^n A'}{A}$, then the differential equation for $s$
is
\begin{equation}
is' - \frac{s^2}{Fr^n} + \frac{\omega^2 r^n}{F} + Y r^n = 0.
\end{equation}
Suppose we write $s$ as a Taylor series about some $\omega_0$ in the
upper half plane:
\begin{equation}
s(r,\omega)=\sum_{k=0}^{\infty} s_k(r) (\omega-\omega_0)^k.
\end{equation}
If the Taylor series has a nonzero radius of convergence,
then $s$ is analytic in a neighborhood of $\omega_0$.
We can write an infinite sequence of differential equations:
\begin{equation} \label{eq:skdiff}
is_k' = \sum_{j=0}^k \frac{s_j s_{k-j}}{Fr^n} - (\delta_{k0} \omega_0^2+2\delta_{k1} \omega_0+\delta_{k2}) \frac{r^n}{F} - Y r^n \delta_{k0}
\end{equation}
From our analysis in Section \ref{sec:contractbosons} and Appendix \ref{app:contractbosons},
we know how $s_0$ behaves.
Note that, for $k>0$, all of the equations are linear in $s_k$.
Since the imaginary part of $\frac{s_0(r_H)}{Fr^n}$ is non-positive, there exists a solution for each $s_k$ that remains finite at the horizon.
We see that the $s_k$ obey a recurrence relation similar to that of
the Catalan numbers.  It is not hard to show using generating functions
that the $s_k$ grow at most exponentially, and thus the radius of
convergence is finite.

In the limit $r \to \infty$, we have to worry about the $s_k$ diverging.
Generically, the terms $s_0,s_1,s_2$ are $O(r^{n-1})$.  Because of the $Fr^n$ in
the denominator of
\eqref{eq:skdiff}, subsequent terms diverge successively more slowly,
and for sufficiently large $k$ there is no divergence.  So we may
renormalize $s$ by subtracting
a function that is polynomial in $\omega$.

Since $\int \frac{1}{Fr^n}$ and $\int \frac{s_k}{Fr^n}$ are bounded as
$r \to \infty$ and all but
finitely many $s_k$ are bounded as $r \to \infty$, our argument
that the $s_k$ grow at most exponentially works in the limit $r \to \infty$ as
well. 



\begin{thebibliography}{JHEP}

\bibitem{Maldacena:1997re}
  J.~M.~Maldacena,
  ``The large N limit of superconformal field theories and supergravity,''
  Adv.\ Theor.\ Math.\ Phys.\  {\bf 2}, 231 (1998)
  [Int.\ J.\ Theor.\ Phys.\  {\bf 38}, 1113 (1999)]
  [arXiv:hep-th/9711200].

\bibitem{Gubser:1998bc}
  S.~S.~Gubser, I.~R.~Klebanov and A.~M.~Polyakov,
  ``Gauge theory correlators from non-critical string theory,''
  Phys.\ Lett.\  B {\bf 428}, 105 (1998)
  [arXiv:hep-th/9802109].

\bibitem{Witten:1998qj}
  E.~Witten,
  ``Anti-de Sitter space and holography,''
  Adv.\ Theor.\ Math.\ Phys.\  {\bf 2}, 253 (1998)
  [arXiv:hep-th/9802150].

\bibitem{Kharzeev:2007wb}
  D.~Kharzeev and K.~Tuchin,
  ``Bulk viscosity of QCD matter near the critical temperature,''
  JHEP {\bf 0809}, 093 (2008)
  [arXiv:0705.4280 [hep-ph]].

\bibitem{Karsch:2007jc}
  F.~Karsch, D.~Kharzeev and K.~Tuchin,
  ``Universal properties of bulk viscosity near the QCD phase transition,''
  Phys.\ Lett.\  B {\bf 663}, 217 (2008)
  [arXiv:0711.0914 [hep-ph]].

\bibitem{Meyer:2010ii}
  H.~B.~Meyer,
  ``The Bulk Channel in Thermal Gauge Theories,''
  JHEP {\bf 1004}, 099 (2010)
  [arXiv:1002.3343 [hep-lat]].

\bibitem{Meyer:2010gu}
  H.~B.~Meyer,
  ``Lattice Gauge Theory Sum Rule for the Shear Channel,''
  arXiv:1005.2686 [hep-lat].

\bibitem{Romatschke:2009ng}
  P.~Romatschke and D.~T.~Son,
  ``Spectral sum rules for the quark-gluon plasma,''
  Phys.\ Rev.\  D {\bf 80}, 065021 (2009)
  [arXiv:0903.3946 [hep-ph]].

\bibitem{Springer:2010mf}
  T.~Springer, C.~Gale, S.~Jeon and S.~H.~Lee,
  ``A shear spectral sum rule in a non-conformal gravity dual,''
  arXiv:1006.4667 [hep-th].

\bibitem{Springer:2010mw}
  T.~Springer, C.~Gale and S.~Jeon,
  ``Bulk spectral functions in single and multi-scalar gravity duals,''
  arXiv:1010.2760 [hep-th].


  \bibitem{Hartnoll:2008vx}
  S.~A.~Hartnoll, C.~P.~Herzog and G.~T.~Horowitz,
  ``Building an AdS/CFT superconductor,''
 Phys.\ Rev.\ Lett.\  {\bf 101}, 031601 (2008)
 [arXiv:0803.3295 [hep-th]].

\bibitem{Baier:2009zy}
  R.~Baier,
  ``R-charge thermodynamical spectral sum rule in N=4 Yang-Mills theory,''
  arXiv:0910.3862 [hep-th].

 \bibitem{Gubser:2008px}
  S.~S.~Gubser,
  ``Breaking an Abelian gauge symmetry near a black hole horizon,''
  arXiv:0801.2977 [hep-th].

 \bibitem{Hartnoll:2008kx}
  S.~A.~Hartnoll, C.~P.~Herzog and G.~T.~Horowitz,
  ``Holographic Superconductors,''
  JHEP {\bf 0812}, 015 (2008)
  [arXiv:0810.1563 [hep-th]].

\bibitem{Amado:2007pv}
  I.~Amado, C.~Hoyos-Badajoz, K.~Landsteiner and S.~Montero,
  ``Absorption Lengths in the Holographic Plasma,''
  JHEP {\bf 0709}, 057 (2007)
  [arXiv:0706.2750 [hep-th]].

\bibitem{Amado:2009ts}
  I.~Amado, M.~Kaminski and K.~Landsteiner,
  ``Hydrodynamics of Holographic Superconductors,''
  JHEP {\bf 0905}, 021 (2009)
  [arXiv:0903.2209 [hep-th]].
  
  \bibitem{Herzog:2008he}
  C.~P.~Herzog, P.~K.~Kovtun and D.~T.~Son,
  ``Holographic model of superfluidity,''
  Phys.\ Rev.\  D {\bf 79}, 066002 (2009)
  [arXiv:0809.4870 [hep-th]].
  
\bibitem{Son:2002sd}
  D.~T.~Son and A.~O.~Starinets,
  ``Minkowski-space correlators in AdS/CFT correspondence: Recipe and
  applications,''
  JHEP {\bf 0209}, 042 (2002)
  [arXiv:hep-th/0205051].

\bibitem{Herzog:2002pc}
  C.~P.~Herzog and D.~T.~Son,
  ``Schwinger-Keldysh propagators from AdS/CFT correspondence,''
  JHEP {\bf 0303}, 046 (2003)
  [arXiv:hep-th/0212072].
  
 \bibitem{Policastro:2002se}
  G.~Policastro, D.~T.~Son and A.~O.~Starinets,
  ``From AdS/CFT correspondence to hydrodynamics,''
  JHEP {\bf 0209}, 043 (2002)
   [arXiv:hep-th/0205052].

\bibitem{Benini:2010pr}
  F.~Benini, C.~P.~Herzog, R.~Rahman and A.~Yarom,
  ``Gauge gravity duality for d-wave superconductors: prospects and
  challenges,''
  arXiv:1007.1981 [hep-th].
 
\bibitem{Kaminski:2009ce}
  M.~Kaminski, K.~Landsteiner, F.~Pena-Benitez, J.~Erdmenger, C.~Greubel and P.~Kerner,
  ``Quasinormal modes of massive charged flavor branes,''
  JHEP {\bf 1003}, 117 (2010)
  [arXiv:0911.3544 [hep-th]].

\bibitem{Gubser:2008wz}
  S.~S.~Gubser and F.~D.~Rocha,
  ``The gravity dual to a quantum critical point with spontaneous symmetry
  breaking,''
  Phys.\ Rev.\ Lett.\  {\bf 102}, 061601 (2009)
  [arXiv:0807.1737 [hep-th]].
  
\bibitem{Hoyos:2006gb}
  C.~Hoyos-Badajoz, K.~Landsteiner and S.~Montero,
  ``Holographic Meson Melting,''
  JHEP {\bf 0704}, 031 (2007)
  [arXiv:hep-th/0612169].
  
\bibitem{Kruczenski:2003be}
  M.~Kruczenski, D.~Mateos, R.~C.~Myers and D.~J.~Winters,
  ``Meson spectroscopy in AdS/CFT with flavour,''
  JHEP {\bf 0307}, 049 (2003)
  [arXiv:hep-th/0304032].

\bibitem{Babington:2003vm}
  J.~Babington, J.~Erdmenger, N.~J.~Evans, Z.~Guralnik and I.~Kirsch,
  ``Chiral symmetry breaking and pions in non-supersymmetric gauge /  gravity
  duals,''
  Phys.\ Rev.\  D {\bf 69}, 066007 (2004)
  [arXiv:hep-th/0306018].

\bibitem{Kirsch:2004km}
  I.~Kirsch,
  ``Generalizations of the AdS/CFT correspondence,''
  Fortsch.\ Phys.\  {\bf 52}, 727 (2004)
  [arXiv:hep-th/0406274].

\bibitem{Mateos:2006nu}
  D.~Mateos, R.~C.~Myers and R.~M.~Thomson,
  ``Holographic phase transitions with fundamental matter,''
  Phys.\ Rev.\ Lett.\  {\bf 97}, 091601 (2006)
  [arXiv:hep-th/0605046].

\bibitem{Horowitz:2008bn}
  G.~T.~Horowitz and M.~M.~Roberts,
  ``Holographic Superconductors with Various Condensates,''
  Phys.\ Rev.\  D {\bf 78}, 126008 (2008)
  [arXiv:0810.1077 [hep-th]].
  
\bibitem{Erdmenger:2007ja}
  J.~Erdmenger, M.~Kaminski and F.~Rust,
  ``Holographic vector mesons from spectral functions at finite baryon or
  isospin density,''
  Phys.\ Rev.\  D {\bf 77}, 046005 (2008)
  [arXiv:0710.0334 [hep-th]].
  
\bibitem{Myers:2008cj}
  R.~C.~Myers and A.~Sinha,
  ``The fast life of holographic mesons,''
  JHEP {\bf 0806}, 052 (2008)
  [arXiv:0804.2168 [hep-th]].

\bibitem{Kaminski:2008ai}
  M.~Kaminski,
  ``Holographic quark gluon plasma with flavor,''
  Fortsch.\ Phys.\  {\bf 57}, 3 (2009)
  [arXiv:0808.1114 [hep-th]].

\bibitem{Faulkner:2009am}
  T.~Faulkner, G.~T.~Horowitz, J.~McGreevy, M.~M.~Roberts and D.~Vegh,
  ``Photoemission 'experiments' on holographic superconductors,''
  JHEP {\bf 1003}, 121 (2010)
  [arXiv:0911.3402 [hep-th]].

\bibitem{Minwalla:1997ka}
  S.~Minwalla,
  ``Restrictions imposed by superconformal invariance on quantum field
  theories,''
  Adv.\ Theor.\ Math.\ Phys.\  {\bf 2}, 781 (1998)
  [arXiv:hep-th/9712074].

\bibitem{Iqbal:2009fd}
  N.~Iqbal and H.~Liu,
  ``Real-time response in AdS/CFT with application to spinors,''
  Fortsch.\ Phys.\  {\bf 57}, 367 (2009)
  [arXiv:0903.2596 [hep-th]].

\bibitem{Liu:2009dm}
  H.~Liu, J.~McGreevy and D.~Vegh,
  ``Non-Fermi liquids from holography,''
  arXiv:0903.2477 [hep-th].
  
\bibitem{Faulkner:2009wj}
  T.~Faulkner, H.~Liu, J.~McGreevy and D.~Vegh,
  ``Emergent quantum criticality, Fermi surfaces, and AdS2,''
  arXiv:0907.2694 [hep-th].

\bibitem{Cubrovic:2009ye}
  M.~Cubrovic, J.~Zaanen and K.~Schalm,
  ``String Theory, Quantum Phase Transitions and the Emergent Fermi-Liquid,''
  Science {\bf 325}, 439 (2009)
  [arXiv:0904.1993 [hep-th]].

\bibitem{Kaminski:2009dh}
  M.~Kaminski, K.~Landsteiner, J.~Mas, J.~P.~Shock and J.~Tarrio,
  ``Holographic Operator Mixing and Quasinormal Modes on the Brane,''
  JHEP {\bf 1002}, 021 (2010)
  [arXiv:0911.3610 [hep-th]].

\bibitem{Hartman:2010fk}
  T.~Hartman and S.~A.~Hartnoll,
  ``Cooper pairing near charged black holes,''
  JHEP {\bf 1006}, 005 (2010)
  [arXiv:1003.1918 [hep-th]].

\bibitem{Gubser:2010dm}
  S.~S.~Gubser, F.~D.~Rocha and A.~Yarom,
  ``Fermion correlators in non-abelian holographic superconductors,''
  arXiv:1002.4416 [hep-th].

\bibitem{Gubser:2008ny}
  S.~S.~Gubser and A.~Nellore,
  ``Mimicking the QCD equation of state with a dual black hole,''
  Phys.\ Rev.\  D {\bf 78}, 086007 (2008)
  [arXiv:0804.0434 [hep-th]].
  
\bibitem{Kanitscheider:2009as}
  I.~Kanitscheider and K.~Skenderis,
  ``Universal hydrodynamics of non-conformal branes,''
  JHEP {\bf 0904}, 062 (2009)
  [arXiv:0901.1487 [hep-th]].

\bibitem{Hartnoll:2005ju}
  S.~A.~Hartnoll and S.~Prem Kumar,
  ``AdS black holes and thermal Yang-Mills correlators,''
  JHEP {\bf 0512}, 036 (2005)
  [arXiv:hep-th/0508092].
  
\bibitem{Kovtun:2003vj}
  P.~Kovtun and L.~G.~Yaffe,
  ``Hydrodynamic fluctuations, long-time tails, and supersymmetry,''
  Phys.\ Rev.\  D {\bf 68}, 025007 (2003)
  [arXiv:hep-th/0303010].

\bibitem{Buchel:2010gd}
  A.~Buchel,
  ``Critical phenomena in N=4 SYM plasma,''
  Nucl.\ Phys.\  B {\bf 841}, 59 (2010)
  [arXiv:1005.0819 [hep-th]].

\bibitem{Damascelli:2003bi}
  A.~Damascelli, Z.~Hussain and Z.~X.~Shen,
  Rev.\ Mod.\ Phys.\  {\bf 75}, 473 (2003).

\bibitem{Kovtun:2005ev}
  P.~K.~Kovtun and A.~O.~Starinets,
  ``Quasinormal modes and holography,''
  Phys.\ Rev.\  D {\bf 72}, 086009 (2005)
  [arXiv:hep-th/0506184].
  
\bibitem{Ren:2010ha}
  J.~Ren,
  ``One-dimensional holographic superconductor from $AdS_3/CFT_2$
  correspondence,''
  arXiv:1008.3904 [hep-th].
  
\bibitem{Birmingham:2001pj}
  D.~Birmingham, I.~Sachs and S.~N.~Solodukhin,
  ``Conformal field theory interpretation of black hole quasi-normal modes,''
  Phys.\ Rev.\ Lett.\  {\bf 88}, 151301 (2002)
  [arXiv:hep-th/0112055].

\bibitem{Herzog:2007ij}
  C.~P.~Herzog, P.~Kovtun, S.~Sachdev and D.~T.~Son,
  ``Quantum critical transport, duality, and M-theory,''
  Phys.\ Rev.\  D {\bf 75}, 085020 (2007)
  [arXiv:hep-th/0701036].
  
\bibitem{Myers:2007we}
  R.~C.~Myers, A.~O.~Starinets and R.~M.~Thomson,
  ``Holographic spectral functions and diffusion constants for fundamental
  matter,''
  JHEP {\bf 0711}, 091 (2007)
  [arXiv:0706.0162 [hep-th]].

\end{thebibliography}
\end{document}